\PassOptionsToPackage{svgnames,table,xcdraw}{xcolor}
\documentclass[journal, preprint]{vgtc_arxiv} 
\usepackage{graphicx}
\usepackage{times}
 
\usepackage{mathptmx}                  
\DeclareMathAlphabet{\mathcal}{OMS}{cmsy}{m}{n} 
\usepackage{amsmath}
\usepackage{amssymb}
\usepackage{booktabs}
\usepackage{placeins}
\usepackage[pagebackref=true,breaklinks=true, colorlinks,bookmarks=false,pagebackref,bookmarks]{hyperref}

\preprinttext{}

\vgtccategory{Research}

\usepackage[ruled]{algorithm2e} 

\SetAlFnt{\small}
\SetAlCapFnt{\small}
\SetAlCapNameFnt{\small}
\SetAlCapHSkip{0pt}

\newcommand{\red}[1]{\cellcolor[HTML]{f08484}{#1}}
\newcommand{\green}[1]{\cellcolor[HTML]{8cdeb2}{#1}}
\newcommand{\yellow}[1]{\cellcolor[HTML]{ffba49}{#1}}

\newcommand{\etal}[1]{et al. }
\newcommand{\Ip}{\mathbf{I}_\text{p}}
\newcommand{\Ic}{\mathbf{I}_\text{c}}
\newcommand{\xp}{\mathbf{x}_\text{p}}
\newcommand{\xc}{\mathbf{x}_\text{c}}
\newcommand{\xs}{\mathbf{x}_\text{s}}
\newcommand{\Ep}{\mathbf{E}_\text{p}}
\newcommand{\Ec}{\mathbf{E}_\text{c}}
\newcommand{\Lp}{\mathbf{L}_\text{p}}
\newcommand{\Li}{\mathbf{L}_\text{i}}

\newcommand{\Kp}{\mathbf{K}_\text{p}}
\newcommand{\Kc}{\mathbf{K}_\text{c}}
\newcommand{\Gp}{g_\text{p}}
\newcommand{\Gc}{g_\text{c}}

\newcommand{\gp}{\boldsymbol{\gamma}_\text{p}}
\newcommand{\Vn}{\mathbf{v}_\text{n}}
\newcommand{\gc}{\boldsymbol{\gamma}_\text{c}}
\newcommand{\R}{\mathbf{r}}

\newcommand{\w}{\mathbf{w}}

\usepackage{tabularx}
\usepackage[nameinlink,capitalise]{cleveref}
\usepackage{svg}

\crefname{figure}{Fig.}{Figs.}
\crefname{equation}{Eq.}{Eqs.}
\crefname{table}{Table}{Tables}
\usepackage{float}

\title{DPCS: Path Tracing-Based Differentiable Projector-Camera Systems}
\supptitle{DPCS: Path Tracing-Based Differentiable Projector-Camera Systems}
\author{
  \authororcid{Jijiang Li}{0009-0003-2823-751X},
  \authororcid{Qingyue Deng}{0009-0001-8471-2480},
  \authororcid{Haibin Ling}{0000-0003-4094-8413}, and 
  \authororcid{Bingyao Huang}{0000-0002-8647-5730}
}

\authorfooter{
  \item
  	Jijiang Li is with Southwest University.
  	E-mail: jijiangli@email.swu.edu.cn.
  \item
  	Qingyue Deng is with Southwest University.
  	E-mail: qingyue@email.swu.edu.cn.
  \item
  	Haibin Ling is with Stony Brook University.
  	E-mail: hling@cs.stonybrook.edu.

  \item Bingyao Huang is with Southwest University.
  	E-mail: bhuang@swu.edu.cn. Corresponding author.
}

\abstract{
Projector-camera systems (ProCams) simulation aims to model the physical project-and-capture process and associated scene parameters of a ProCams, and is crucial for spatial augmented reality (SAR) applications such as ProCams relighting and projector compensation. Recent advances use an end-to-end neural network to learn the project-and-capture process. However, these neural network-based methods often implicitly encapsulate scene parameters, such as surface material, gamma,  and white balance in the network parameters, and are less interpretable and hard for novel scene simulation. 
Moreover, neural networks usually learn the indirect illumination implicitly in an image-to-image translation way which leads to poor performance in simulating complex projection effects such as soft-shadow and interreflection.
In this paper, we introduce a novel path tracing-based differentiable projector-camera systems (DPCS), offering a differentiable ProCams simulation method that explicitly integrates multi-bounce path tracing. Our DPCS models the physical project-and-capture process using differentiable physically-based rendering (PBR), enabling the scene parameters to be explicitly decoupled and learned using much fewer samples. Moreover, our physically-based method not only enables high-quality downstream ProCams tasks, such as ProCams relighting and projector compensation, but also allows novel scene simulation using the learned scene parameters.
In experiments, DPCS demonstrates clear advantages over previous approaches in ProCams simulation, offering better interpretability, more efficient handling of complex interreflection and shadow, and requiring fewer training samples.
The code and dataset are available on the project page: \url{https://jijiangli.github.io/DPCS/}.
}


\teaser{
  \includegraphics[width=1\textwidth]{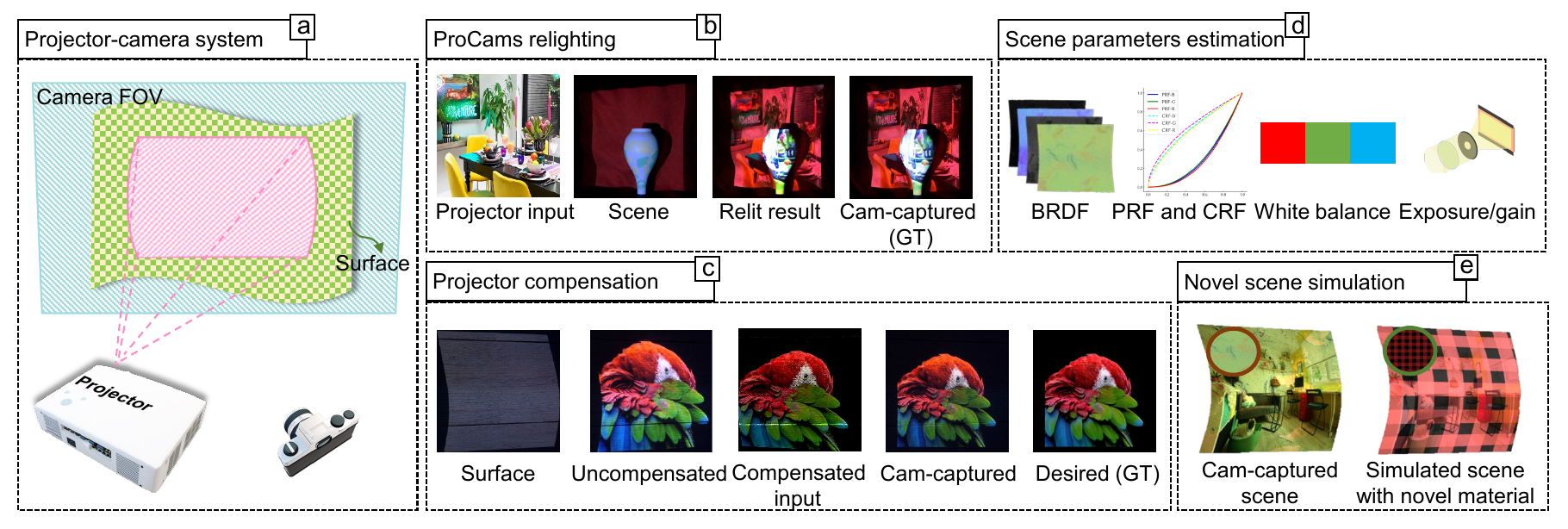}
  \caption{\textbf{DPCS: Path Tracing-Based Differentiable Projector-Camera Systems.} DPCS simulates the physical processes of \textbf{(a) projector-camera systems (ProCams)} which involve projecting light onto the surface, (inter)reflecting light off the surface, and finally capturing the scene by a camera. Once trained, DPCS can be applied to various tasks: for \textbf{(b) relighting}, we can predict the images captured by the camera under the illumination of a novel projection image; for \textbf{(c) projector compensation}, it can be accomplished by differentiating the projector input image to find the optimal compensated projector input image that achieves the desired camera-captured effect; for \textbf{(e) novel scene simulation}, it can be performed by editing \textbf{(d)}  the estimated scene parameters, such as BRDF, projector/camera response functions (PRF/CRF), \textit{etc}.}
  \label{fig:teaser}
}

\begin{document}

\firstsection{Introduction}
\maketitle
Projector-camera systems (ProCams) simulation has many applications in spatial augmented reality (SAR)/projection mapping\cite{grundhofer2018recent,Erel2024CasperDPM,iwai_shadow_mapping,watanabe_face_mapping,watanabe_sphere_mapping,Iwai_env_mapping,Kageyama2024dynamicmapping,IwaimappingLight2023,Ishikawa2024mixmapping,shadow_mapping_iwai_2024}, such as ProCams relighting\cite{deprocams,guo2019relightables,han2014fast,oya2017image,peers2009compressive,ren2015image,sen2005dual,siegl2015real,sueishi2015robust}, projector compensation
\cite{wang2024vicompen,deprocams,compennest_pp,tps,li2023physicsbased, narita2016dynamic,aliaga2012fast,Ashdown2006compensation} and novel scene simulation \cite{nepmap,deprocams}. An example is shown in \cref{fig:teaser} where ProCams relighting, projector compensation, and novel scene simulation can be performed using the proposed differentiable projector-camera systems (DPCS).

Traditional methods per pixel simulate ProCams to accomplish these tasks by calibrating the color mixing properties in ProCams \cite{grossberg2004making,chen2008color}. These methods involve projecting images at varying brightness levels and then capturing the samplings for radiometric calibration, often lacking flexibility and potentially introducing biases.
Other works utilize the light transport matrix (LTM) and aim to establish a linear mapping between the camera-captured images and the projected images using a large matrix \cite{chiba2018ultra,o2010optical,o20143d,sen2005dual,LTM_jiaping, Ren2003relighting}. They also need to perform an additional radiometric calibration. Furthermore, scene parameters such as geometry, material, and lighting are coupled in LTM, making it challenging to decompose these parameters. Recently, a neural rendering framework~\cite{deprocams} was developed to simulate the project-and-capture process, offering impressive results in photorealistic rendering. Despite its efficacy, such a neural network-based method cannot fully decompose scene parameters, such as BRDF and projector/camera radiometric response functions (PRF/CRF). Although it requires a large number of training samples, typically more than 50 projected and captured sampling image pairs, to achieve satisfactory results, the lack of parameter interpretability may lead to suboptimal results in novel scene simulation. Additionally, it implicitly learns the rich direct and indirect light interactions within the network parameters, and a pair of projected and captured gray sampling images is needed as a prior for the reflectance. This may result in poor performance when simulating intricate light transport effects, such as soft shadows and interreflection, especially due to inadequate indirect illumination priors in the camera-captured surface image, which is approximated by a scene image captured under gray illumination.

In this paper, we show that it is possible to simulate ProCams using a physically-based differentiable rendering approach, \textit{i.e.}, DPCS rather than a neural network. In particular, we model the physical project-and-capture process as a physically-based rendering (PBR) problem that explicitly renders the camera-captured projection using differentiable path tracing. We start by modeling the projector as an area light that accepts sRGB images as input. The input images are first processed by the projector radiometric transfer function: 3-channel gammas and gain. Then, the modulated projector light shines on the textured projection surface, where light is (inter)reflected according to the surface mesh and BRDF. Finally, the reflected surface light paths illuminate the camera sensor and the irradiance is converted back to sRGB according to the camera response function: exposure (gain) and 3-channel gammas. To solve the unknown scene parameters of DPCS, we first project and capture a few images, and use the projected and captured image pairs to optimize the scene parameters using inverse rendering in a self-supervised manner, such that the optimal scene parameters may align the rendered camera-captured images with the real camera-captured ones.
After a few minutes of data capturing and training, our DPCS is ready to address downstream SAR tasks, \textit{e.g.}, projector compensation and ProCams relighting. Furthermore, the explicit simulation of ProCams through physically-based rendering enables straightforward editing of scene parameters, such as geometry and surface materials, allowing novel scene simulation.

\begin{table*}[t]
    \caption{Comparison of representative existing methods and our DPCS.}
    \label{table:proj_rel_work}
    \centering

    \resizebox{\linewidth}{!}{
    \begin{tabular}{|l|c|c|c|c|c|c|}
    \hline
     Method& ProCams & Projector& Geometry & PBR BRDF & CRF/PRF & Novel scene \\
 & relighting& compensation& editing& estimation & estimation & simulation\\
\hline
    Matrix-based~\cite{grossberg2004making} & \green{Yes} & \green{Yes} & \red{No} & \red{No}
    & \red{No}
     &\red{No}
\\
     Matrix-based + optical flow
    ~\cite{li2023physicsbased}& \yellow{Simple}& \yellow{Simple} & \red{No} & \red{No}
& \green{Yes} &     \yellow{CRF/PRF, BRDF}\\
    Rough shading + CNN~\cite{deprocams} & \green{Yes} & \green{Yes} & \yellow{Partial}  &     \red{No}
& \red{No} &\yellow{Pose, geometry}\\
    NeRF-based~\cite{nepmap} & \green{Yes}& \green{Yes} & \red{No}  &     \green{Yes} &\green{Yes} &\yellow{CRF/PRF, pose}\\
    DR-based + iterative refine~\cite{diff_projector}  & \red{No}& \green{Yes} & \green{Yes}  &     \red{No} &\red{No} &\yellow{Pose, geometry}\\
    DPCS (ours)& \green{Yes} & \green{Yes} & \green{Yes}  &     \green{Yes} &\green{Yes} &\green{CRF/PRF, pose, BRDF, geometry}\\
    \hline
    \end{tabular}
    }
    \begin{flushleft}
    \textbf{Abbreviations:} (CNN) Convolutional Neural Network, (DR) Differentiable Rendering.
    \end{flushleft}
    
\end{table*}

Our contribution can be summarized as follows:
\begin{itemize}
    \vspace{-1mm}\item A novel differentiable simulation method for ProCams named DPCS which integrates multi-bounce path tracing, enabling high-quality simulation of complex light interactions in ProCams. We demonstrate the efficiency in downstream ProCams tasks such as ProCams relighting, projector compensation, and novel scene simulation using DPCS.
    \item DPCS can decompose ProCams as physically-based differentiable parameters such as camera pose, nonlinear responses, surface materials, and white balance coefficients. These parameters can be edited to perform novel scene simulation.
    \item DPCS operates efficiently without the need for excessive samplings for training.
\end{itemize}

\vspace{-1mm}For evaluation, we applied DPCS to tasks including ProCams relighting, projector compensation, and novel scene simulation. The effectiveness of DPCS is clearly demonstrated in the experiments compared with existing solutions.

\section{Related Work}

\subsection{ProCams simulations}
Classical ProCams simulation methods use LTM
\cite{o2010optical,o20143d,sen2005dual, LTM_jiaping,Ren2003relighting} to model the relationship between the intensity of each camera pixel and the intensities of all projector pixels through linear combinations. Although these methods are effective for accurate ProCams simulation, they often require radiometric calibration or additional devices, such as a second camera \cite{LTM_jiaping} and beamsplitters \cite{o2010optical}. Recently, neural network-based solutions have been proposed to address these issues. For example, DeProCams \cite{deprocams} learns the photometric and geometric mappings between projector input images and camera-captured images without radiometric calibration or additional devices. However, such a neural rendering framework may not fully adhere to physical imaging principles, and some scene parameters are coupled in the network parameters, which remain uninterpretable. For the first time, Erel \etal~\cite{nepmap} implement multi-view projection mapping through neural reflectance fields \cite{srinivasan2021nerv,zhang2021nerfactor} which can synthesize novel viewpoint images of the ProCams. However, the NeRF-like approach optimizes several Multi-Layer Perceptrons (MLPs) using volumetric ray-marching, which is both time and memory intensive during training and inference. Additionally, implicitly representing geometry, BRDF, and transmittance using MLPs hinders editing scene parameters for novel scene simulations. The above methods aim to simulate the physical project-and-capture process in ProCams, facilitating various important SAR tasks that rely on accurate ProCams simulations.

\textbf{ProCams relighting} aims to synthesize photorealistic images that replicate the effect of projecting new light patterns onto a scene, as if a real camera would capture them. This approach is particularly valuable in spatial augmented reality (SAR) and projection mapping, enabling the design, testing, and debugging of new light patterns in a virtual scene before actual projection, allowing for edits and adjustments to optimize the projection mapping effects. Traditional methods involve either fitting a nonlinear color mapping function for each pixel \cite{grossberg2004making,tps}, or computing an LTM \cite{chiba2018ultra,o2010optical,o20143d,sen2005dual, LTM_jiaping,Ren2003relighting} from projected and captured sampling image pairs.
These precomputed models can subsequently be applied to novel projection images for predicting corresponding camera-captured images. Recently, neural rendering approaches \cite{deprocams, nepmap} learn the shading properties from training samples and infer the camera-captured images without additional radiometric calibration. 
Unfortunately, neural network-based methods \cite{compennest_pp,deprocams} implicitly model complex light interactions within network parameters and still need a gray projection sampling image to obtain direct and indirect reflectance as input prior, which may struggle to realistically simulate the complex interreflection of projection illumination within a scene. \cref{fig:relighting} shows ProCams relighting results of different methods. This motivates us to design a method that addresses this by explicitly simulating light path interactions within the ProCams scene.

\textbf{Projector compensation} aims to modify the projector input image to cancel the geometric and photometric distortions from the environment and projection surfaces, and to improve the viewer's perception experience.
A classical solution~\cite{grossberg2004making,tps} models the nonlinear image formation process between the projector and camera-captured images per pixel and then inverts this process to obtain the compensated projector input. Recent methods leverage deep learning \cite{deprocams,compennest_pp,li2023physicsbased, Wang2023compenHR, Huang2019CompenNet++} to learn image-to-image mapping, achieving impressive results. However, such implicit representations may lack interpretability and often require a large number of training samples. The differentiable rendering (DR) + iterative refinement method \cite{diff_projector} defines projector compensation as a differentiable rendering problem. However, due to this approach's limitations in modeling the BRDF and nonlinear response, it may not estimate projection surface materials or CRF/PRF, making it hard to perform ProCams relighting and other SAR tasks. Moreover, their differentiable compensation framework necessitates the use of an extra thin plate spline-based pixel shifting technique \cite{TPS_spline} and includes an iterative refinement procedure \cite{nayar2003projection}. This procedure involves repeatedly projecting and capturing several hundred samples to estimate bias and calibrate the projector input. Consequently, this approach is both challenging to adapt to different SAR tasks and time-intensive, requiring not only offline training for compensated inputs but also further iterative refinements.
The matrix-based and optical flow approach outlined in \cite{li2023physicsbased} employs a color mixing matrix for material modeling and leverages optical flow for geometric corrections, achieving rapid results in both training and inference. However, it assumes that projection surfaces contain only a limited number of simple signatures, allowing materials to be represented as a linear combination of a few basis functions. This assumption poses challenges for more complex textured scenes, potentially leading to instability when estimating scene parameters.

\subsection{Inverse rendering}
Inverse rendering aims to determine scene parameters given observed images \cite{azinovic2019inverse,tsai2019beyond}. Recently, the development of general-purpose tools such as Redner \cite{li2018differentiable}, Mitsuba 2/3 \cite{NimierDavidVicini2019Mitsuba2}, and PSDR-CUDA \cite{PSDR-CUDA} have advanced the field of inverse rendering. Several methods were proposed to estimate the gradients with respect to geometry \cite{nicolet2021large, Reparams} and some other studies \cite{Cai:2022:PSDR-MeshSDF, neuralpbir, NIPS_22_nvidia} estimate both shape and material from a set of captured images. ProCams relighting and projector compensation both incorporate similar concepts of inverse rendering; however, they address more complex lighting conditions, namely projectors containing millions of tiny light sources. Directly implementing the aforementioned differentiable rendering frameworks for ProCams compensation and relighting might not yield optimal outcomes.

\subsection{Our method}
Inspired by the studies above, we leverage physically-based differentiable rendering to simulate ProCams image formation, breaking the scene down into differentiable physical parameters such as camera pose, projector input, surface materials and nonlinear responses, which can be edited after training. The features of the aforementioned ProCams techniques shown in \cref{table:proj_rel_work}, none of the combinations of the techniques can achieve practical ProCams simulation with all features supported in the SAR requirement of time and memory.

\begin{figure*}[!t]
    \centering
    \includegraphics[width=\textwidth]{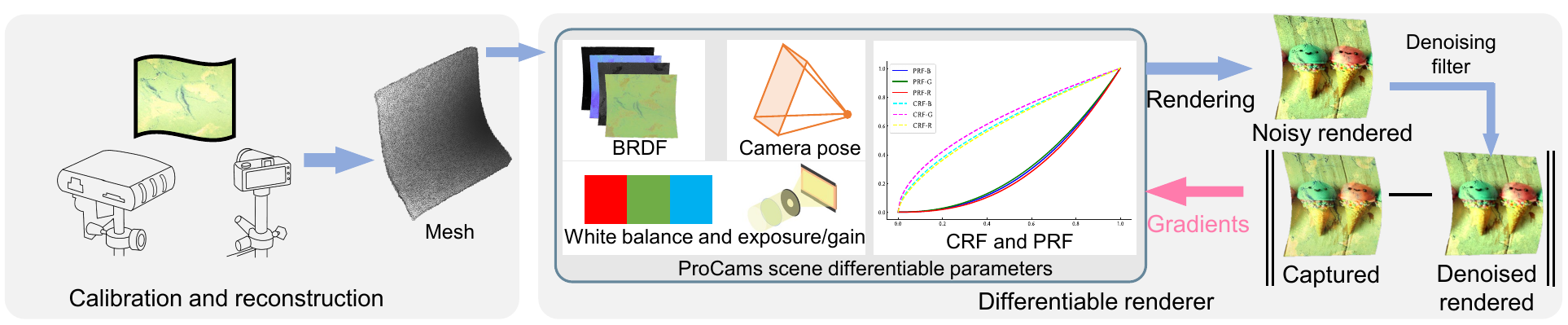}
    \caption{Our physically-based differentiable simulation framework. First, the scene is acquired using structured light (SL) \cite{huang2020fast} to calibrate and reconstruct the surface as a point cloud, which is then utilized to reconstruct the surface into a mesh format. Then, a forward differentiable rendering works to simulate the light transport of the ProCams using predefined scene parameters which contain the surface reflectance, projector response function, and camera response function. Other physical factors like the white balance coefficients, can also be defined for more accurate simulation. The forward rendering approach gives noisy rendered images of the different projection lighting captured by the camera which can be used to calculate pixel loss to the real capturing. Once a denoising filter is applied to the noisy rendered image, it can be leveraged in a gradient-based optimization to minimize the pixel loss between the denoised rendered images and camera-captured images by differentiating the virtual ProCams physical parameters.}
    \label{fig:overview}    
\end{figure*}

\section{Method}
\label{sec:method}

In this paper, we present a differentiable path tracing-based method named DPCS for ProCams simulation. As depicted in \cref{fig:overview}, our goal is to model the physical project-and-capture process (we call it forward rendering) using a physically-based rendering pipeline. Our DPCS incorporates multiple scene components: the projector that functions as the light source (emitter), the projection surface, the camera, and the surface material that affects how light interacts with the surface. We further expand on the system's radiometric responses, which are crucial for accurately modeling the complex responses of the camera and projector. These responses are typically characterized by gamma curves, gain, and white balance. Once the model is trained, we can solve the projector input and scene parameters given a camera-captured image, and we call this process inverse rendering.

\begin{figure}[t]
    \centering
    \includegraphics[width=1\columnwidth]{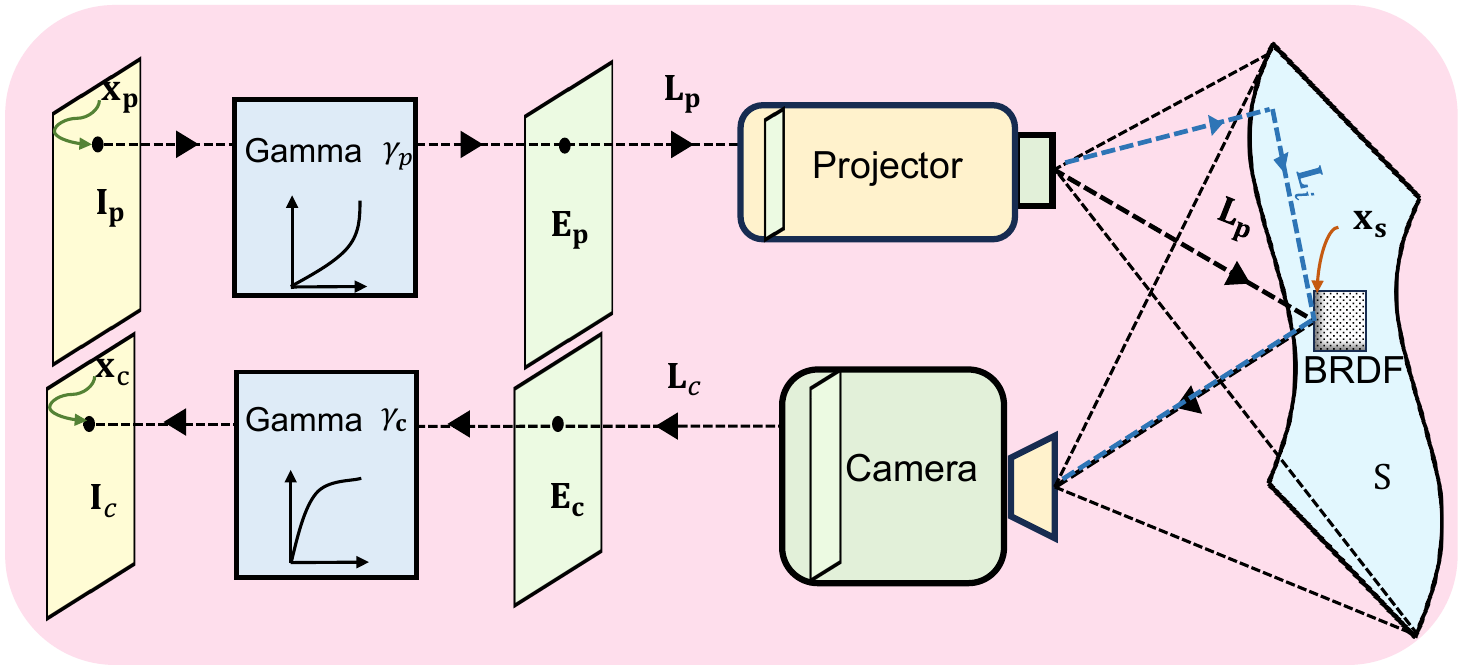}
    \caption{\textbf{ProCams imaging process.} We assume that the scene contains only a projected light source. The scene is illuminated by the direct lighting $ \Lp $ emitted by the projected light source and the indirect lighting $\Li$ resulting from multiple reflections, which is ultimately captured by the camera. The nonlinear transformations of the camera and projector are expressed using gamma functions.}
    \label{fig:method}
\end{figure}
\subsection{Problem formulation}
\label{sec:formulation}
We model the physical project-and-capture process of ProCams as a mapping from the projector input image $ \Ip $ to the camera-captured image $ \Ic $. 
Our method outlines the forward simulation of ProCams, as illustrated in \cref{fig:method}. We assume that the world origin is at the camera optical center. Given a pixel $ \xp \in \mathbb{R}^{2} $ in the projector image $ \Ip $, its emitted irradiance $\Ep(\xp,j)$ is given by:
\begin{equation}
    \Ep(\xp, j)=\Gp\ \Ip(\xp, j)^{\gp(j)},
    \label{eq:PRF}
\end{equation}
where $j$ is the RGB channel, $\gp(j)$ is the projector's gamma of the channel $j$, and $\Gp$ is the projector's gain (we assume a uniform gain for all pixels). \cref{eq:PRF} describes the nonlinear transformation from projector input pixel intensity to projected irradiance.

Thus, the projector pixel $\xp$ emitted radiance $\Lp(\xp,j)$ is
\begin{equation}
  \Lp(\xp,j) = \Ep(\xp,j)/\phi,
    \label{eq: light intensity}
\end{equation} 
where $\phi < \pi$ represents the solid angle over which the projector pixel $\xp$ emits light uniformly, and we assume it is a constant for all projector pixels. We assume that the projector is the only light source in the scene. The surface reflected radiance $L_\text{s}(\boldsymbol{\omega}_\text{o},\xs)$ with the direction $\boldsymbol{\omega}_\text{o}$ at any given point in the scene $\xs\in\mathbb{R}^3$ can thus be calculated using the rendering equation \cite{kajiya1986rendering}.

\begin{equation}
 L_\text{s}(\mathbf{\omega_\text{o},\xs}) = \int_{\mathcal{H}} f_\text{r}(\boldsymbol{\omega}_\text{i}, \boldsymbol{\omega}_\text{o},\xs) \left(L_\text{p}(\boldsymbol{\omega}_\text{i},\xs) + L_\text{i}(\boldsymbol{\omega}_\text{i},\xs)\right) (\mathbf{n} \cdot \boldsymbol{\omega}_\text{i}) \, d\boldsymbol{\omega}_\text{i},
\label{eq:radiance intensity}
\end{equation}
where $\mathcal{H}$ is all directions of the unit hemisphere over the surface point $\xs$,  $L_\text{p}(\boldsymbol{\omega}_\text{i},\xs)=\Lp(\xp,j)$ represents the direct radiance from the projector pixel $\xp$ and incident at $\xs$. $L_\text{i}(\boldsymbol{\omega}_\text{i},\xs)$ is the indirect radiance incident at $\xs$. $f_\text{r}(\boldsymbol{\omega}_\text{i}, \boldsymbol{\omega}_\text{o},\xs)$ refers to the surface BRDF. 

Then, the camera-captured scene irradiance $\Ec(\xc,j)$ is 
\begin{equation}
\Ec(\xc,j) =\int_{\mathcal{S}} L_\text{c}(\xc,\xs)  d\xs,
\label{eq: camera irradiance}
\end{equation}
where $L_\text{c}(\xc,\xs)$ is the surface reflected radiance $L_\text{s}(\boldsymbol{\omega},\xs)$ modulated by the camera lens and incident at the camera pixel $\xc$. $\mathcal{S}$ represents the set of all the surfaces of objects in the scene.

Finally, the camera-captured image pixel intensity $\Ic(\xc,j)$ is given by:
\begin{equation}
     \Ic(\xc,j) = \left(\Gc \w(j) \Ec(\xc,j)\right)^{\gc(j)},
    \label{eq:CRF}
\end{equation}
where $\Gc$ is the camera exposure, $\w$ and $\gc$ are the camera white balance coefficients and gamma, respectively. As shown in \cref{fig:method}, the main process involves capturing both indirect and direct radiance from the projector in the scene through (inter)reflection by the camera.

Given the camera and projector intrinsics, $\Kc, \Kp$, rotation matrix $\R_\text{p}$ and translation vector $\mathbf{t}_\text{p}$ between the projector and the camera, we can find the relationship between $\xc,\xp$ and $\xs$ by:
\begin{equation}
    \xc = \Kc\xs,\quad
    \xp = \Kp[\R_\text{p}\mid\mathbf{t}_\text{p}]\xs.    
\label{eq:xc_xp}
\end{equation}

\subsection{Differentiable Projector-Camera Systems (DPCS)}
\label{sec:optimization task}
Our DPCS aims to model the physical project-and-capture process above using differentiable physically-based rendering, so the model can be quickly trained using an analysis-by-synthesis optimization (aka. inverse rendering) and can be applied to downstream SAR tasks simultaneously. 

We start by modeling the projection surface BRDF $f_\textbf{r}$ using Disney principled BRDF \cite{Bur12} and denote it as $\mathbf{m}$. Then, we aggregate \cref{eq:PRF} to \cref{eq:xc_xp}, and define the forward rendering and inverse rendering as
\begin{align}
    &\text{forward rendering}: &\Ic = \mathcal{R}(\Ip, \boldsymbol{\theta}) \label{eq:forward_rendering}\\ 
    &\text{inverse rendering}: &\{\Ip, \boldsymbol{\theta}\} = \mathcal{R}^{-1}(\Ic)
    \label{eq:inverse_rendering}
\end{align}

During training, our goal is to estimate/refine the scene parameters $\boldsymbol{\theta} = \{\mathbf{m}, \w, \gp, \gc, \Vn, \R_\text{c}, \mathbf{t}_\text{c}\}$ using physically-based inverse rendering with a few training samples $\{\Ip^{(i)}, \Ic^{(i)}\}^{K}_{i=1}$ by:
\begin{equation}
    \hat{\boldsymbol{\theta}} = \underset{\boldsymbol{\theta}}{\operatorname{argmin}} \frac{1}{K} \sum_i^K \left| 
    \hat{\mathbf{I}}_\text{c}^{(i)} - \Ic^{(i)}
    \right| + \lambda_{\text{reg}}\mathcal{L}(\mathbf{m}),
    \label{eq:optimization}
\end{equation}
where $\R_\text{c},\mathbf{t}_\text{c}$ represent the rotation matrix and translation vector of the camera to perform minor refinement. $\Vn$ is the normal map in which values are specified relative to the surface normal, following Mitsuba; i.e., a value of $(0, 0, 1)$ in the normal map causes no change to the surface normal. $\hat{\mathbf{I}}_\text{c}^{(i)}=\mathcal{R}(\Ip^{(i)}, \boldsymbol{\theta})$ is our DPCS rendered camera-captured scene under superimposed projection. $\mathcal{L}(\mathbf{m})$ is a total variance loss for smooth BRDF estimation, and $\lambda_{\text{reg}}$ is its weight.

\vspace{1mm}\noindent\textbf{Variance noise reduction}.
The rendering equation \cref{eq:radiance intensity} can be solved using Monte Carlo integration:
\begin{equation}
 L_\text{s}(\mathbf{\omega_\text{o}},\xs) \approx \frac{1}{N} \sum_{k=1}^{N} \frac{f_\text{r}(\boldsymbol{\omega}_k, \boldsymbol{\omega}_\text{o},\xs) \left(L_\text{p}(\boldsymbol{\omega}_k,\xs) + L_\text{i}(\boldsymbol{\omega}_k,\xs)\right) (\mathbf{n} \cdot \boldsymbol{\omega}_k)}{p(\boldsymbol{\omega}_k)},
\label{eq:montecarlo_radiance_approx}
\end{equation}
where \( \boldsymbol{\omega}_k \) is the sampled direction according to the distribution \( p(\boldsymbol{\omega}_k) \), $N$ is the number of samples, and \( p(\boldsymbol{\omega}_k) \) is the probability density function for sampling the direction \( \boldsymbol{\omega}_k \). Consequently, the variance from such random sampling can introduce local variance in the rendered image $\hat{\mathbf{I}}_\text{c}$, which may result in noisy gradients during optimization steps in differentiable rendering. A common approach to reducing this noise is to increase the sample size per pixel. However, this is often impractical due to the significant increase in computational cost. To address this, we apply a differentiable denoising operator before performing inverse optimization with respect to the scene parameters:
\begin{equation}
    \Ic' = D(\hat{\mathbf{I}}_\text{c}) \label{denoiser},
\end{equation}
where $D$ is a differentiable cross-bilateral filter \cite{cross_denoiser,target_aware} to reduce noise in the rendered image before gradient-based optimization.

\vspace{1mm}\noindent\textbf{Projector/camera response functions and white balance.}
The projector and camera nonlinear radiometric response functions (PRF/CRF) in \cref{eq:CRF,eq:PRF} are modeled by per-channel gamma curves ($\gp$ and $\gc$), gain/exposure ($\Gp, \Gc$) and white balance coefficients ($\w$). The explicit modeling of PRF/CRF is crucial because the projection pattern could be inaccurately baked into the estimated BRDF without proper constraint and decoupling. On the other hand, accurately measuring the real PRF/CRF is cumbersome, and we hope to learn them from the data instead. To mitigate this issue, we model $\w, \gc, \gp$ as $3\times1$  vectors, and set $\Gp, \Gc$ to a constant such that a white projector input pattern slightly overexposes the rendered camera-captured image. The gamma of the projector, $\gp$, is restricted to the interval $[2, 3]$. Meanwhile, the camera response function, $\gc$, is limited to the range $[1/3, 1]$. 
We also constrain the white balance coefficients $\w$ to fall within $(0.2,2.5)$. By clearly defining these constraints, we facilitate the explicit separate estimation of each scene parameter. 

\vspace{1mm}\noindent\textbf{Radiance clipping}.
In practice, while sampling the radiance intensities using Monte Carlo integration \cref{eq:montecarlo_radiance_approx} with multiple bounces, if the HDR range of the scene is not specified, some pixel intensities $\Ic(\xc,j)$ may exceed the maximum range of an LDR (Low Dynamic Range) image. This mainly comes from very low probability paths that contribute intensely to equation \cref{eq:montecarlo_radiance_approx}. To address this, we use radiance clipping operation to prevent some radiance samples from exceeding the maximum pixel intensities. It ensures that the radiance of outgoing rays that are reflected and captured by the camera does not exceed the initial projected radiance value. \label{sec:radiance clipping}

\begin{equation}
\delta  L_\text{s}(\mathbf{\omega_\text{o}},\xs) = \min(\delta  L_\text{s}(\mathbf{\omega_\text{o}},\xs), k/\min(\w)),
\label{HDR constraints}
\end{equation}
where $k$ is the projector intensity linear scale in virtual scene, $\w$ is the white balance coefficients, $\delta  L_\text{s}(\mathbf{\omega_\text{o}},\xs) $ is the radiance difference computed by \cref{eq:montecarlo_radiance_approx}.

\begin{figure*}[!h]
    \centering
    \includegraphics[width=0.99\textwidth]{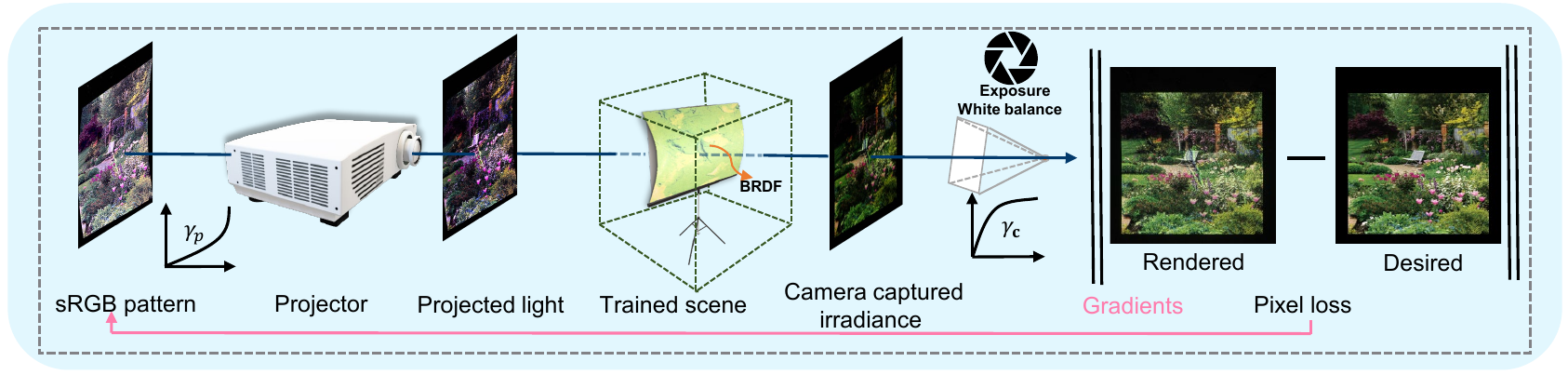}
    \caption{Projector compensation pipeline. Once our DPCS is trained, we can get the compensated projector input image by differentiating the projector input image such that the rendering result is close to the desired appearance.}
    \label{fig:compensation_pipeline}
\end{figure*}

\vspace{1mm}\noindent\textbf{ProCams relighting and projector compensation.}
After training DPCS, we can perform ProCams relighting (i.e., forward rendering) using \cref{eq:forward_rendering}). In particular, given a novel projector input image $\Ip'$, the simulated camera-captured projection $\Ic'$ is given by $\Ic' = \mathcal{R}(\Ip', \boldsymbol{\theta})$.  
Similarly, projector compensation can be performed using \cref{eq:inverse_rendering}: Given a desired viewer-perceived scene under superimpose projection $\Ic$, the process of finding the corresponding projector input image $\Ip^*$ is projector compensation. This can formulated as below and solved via gradient descent, as shown in \cref{fig:compensation_pipeline}:
\begin{align}
    \Ip^* = \mathcal{R}^{-1} (\mathbf{\Ic}) = \underset{\mathbf{\Ip}}{\operatorname{argmin}} \left| 
    \mathcal{R}(\Ip, \boldsymbol{\theta}) - \Ic\right|
    \label{eq:inverse_mapping}
\end{align}

\begin{table}[t!]
    \centering
    \caption{Quantitative comparison on ProCams relighting. Results are averaged over 14 different setups. Clearly, our DPCS methods outperform state-of-the-art CompenNeSt++ \cite{compennest_pp} and DeProCams \cite{deprocams} by a significant margin. This advantage is more notable when the number of training images is small. Compared with two degraded versions of DPCS, despite less than a 1\% difference in PSNR, SSIM, and LPIPS, DPCS exhibits superior visual quality, with reduced local variance and noise compared to \textit{w/o denoiser}. Moreover, DPCS consistently outperforms \textit{w/o clipping} by removing “firefly” in the rendered images. See \cref{fig:ablation study} for a qualitative comparison with its degraded versions. 
    Note that DPCS requires $42$ additional structured light (SL) images for geometric pre-calibration. These samples are not involved in training and are only used to obtain point clouds.}
    \label{tab:compare_relit}
    \resizebox{\columnwidth}{!}{
    \begin{tabular}{@{}c@{\hskip 5pt}c@{\hskip 5pt}l@{\hskip 5pt}c@{\hskip 5pt}c@{\hskip 5pt}c@{\hskip 5pt}c@{}}
    \toprule
        \textbf{\# Train} & \textbf{\# SL} & \textbf{Model} & \textbf{PSNR}$\uparrow$ & \textbf{SSIM}$\uparrow$ & \textbf{LPIPS}$\downarrow$ & $\Delta$\textbf{E}$\downarrow$ \\
        \midrule
        100 & 0  & CompenNeSt++ & 26.3059 & 0.9150 & 0.0928 & 2.2557 \\
            & 0  & DeProCams & 31.6641 & 0.9461 & 0.0625 & 1.2949 \\
            & 42 & DPCS (ours) & 31.7418 & 0.9619 & 0.0322 & \textbf{1.2834} \\
            & 42 & w/o denoiser (ours) & \textbf{31.7798} & \textbf{0.9632} & \textbf{0.0245} & 1.2916 \\
            & 42 & w/o clipping (ours) & 30.8237 & 0.9504 & 0.0529 & 1.3410 \\
        \midrule
        50 & 0  & CompenNeSt++ & 25.6738 & 0.9114 & 0.0962 & 2.3979 \\
           & 0  & DeProCams & 31.4975 & 0.9454 & 0.0630 & 1.3025 \\
           & 42 & DPCS (ours) & \textbf{31.7125} & 0.9616 & 0.0324 & \textbf{1.2785} \\
           & 42 & w/o denoiser (ours) & 31.7019 & \textbf{0.9624} & \textbf{0.0243} & 1.3008 \\
           & 42 & w/o clipping (ours) & 30.8792 & 0.9505 & 0.0523 & 1.3427 \\
        \midrule
        15 & 0  & CompenNeSt++ & 24.3686 & 0.8764 & 0.1356 & 2.9804 \\
           & 0  & DeProCams & 29.6871 & 0.9317 & 0.0748 & 1.5806 \\
           & 42 & DPCS (ours) & 31.7486 & 0.9604 & 0.0326 & \textbf{1.3344} \\
           & 42 & w/o denoiser (ours) & \textbf{31.7687} & \textbf{0.9611} & \textbf{0.0253} & 1.3538 \\
           & 42 & w/o clipping (ours) & 30.8156 & 0.9485 & 0.0540 & 1.4032 \\
        \midrule
        5 & 0  & CompenNeSt++ & 19.9104 & 0.8175 & 0.1873 & 4.6842 \\
          & 0  & DeProCams & 23.6190 & 0.8644 & 0.1456 & 3.3056 \\
          & 42 & DPCS (ours) & 30.9468 & 0.9540 & 0.0355 & 1.5344 \\
          & 42 & w/o denoiser (ours) & \textbf{31.0272} & \textbf{0.9551} & \textbf{0.0295} & \textbf{1.5231} \\
          & 42 & w/o clipping (ours) & 30.0819 & 0.9424 & 0.0572 & 1.6045 \\
    \bottomrule
    \end{tabular}}
\end{table}

\begin{table}[t]
    \centering
    \caption{Resource usage comparison. The reconstruction time includes additional operations such as point cloud reconstruction, mesh generation, and UV unwrapping, as required by DPCS. DPCS has longer inference times for large samples per pixel (SPP) because we explicitly model global illumination with path tracing and perform extra scene parameters (BRDF, CFR/PRF, white balance, etc.) estimation. Note that DPCS  maintains competitive training times. Moreover, its GPU memory usage offers significant advantages over neural network-based methods, making it more suitable for high-resolution ProCams applications.} \label{tab:compare_time_GPU}
    \setlength{\tabcolsep}{4pt}
    \begin{tabular}{@{}l l l c c@{}}
    \toprule
        \textbf{Model} & Training & Memory& Inference & Reconstruction\\
        \multicolumn{1}{c}{} & \multicolumn{1}{c}{time (s)} & \multicolumn{1}{c}{(MB)} & \multicolumn{1}{c}{time (s)} & \multicolumn{1}{c}{time (s)} \\
        \midrule
         DeProCams & 265.02& 	13,904&	 0.0066&\textbf{0}\\
           CompenNeSt++& 1,312.88& 25,289& \textbf{0.0048}&\textbf{0}\\
          \midrule
         DPCS (SPP = 16)& \textbf{218.78}&\textbf{7,098}& 	2.4808& \textasciitilde
 20\\
        DPCS (SPP = 35)& 419.28&  7,103&2.6224& \textasciitilde
 20\\ 
 \midrule
    \end{tabular}
\end{table}

\begin{figure*}[h!]
    \centering
    \includegraphics[width=\textwidth]{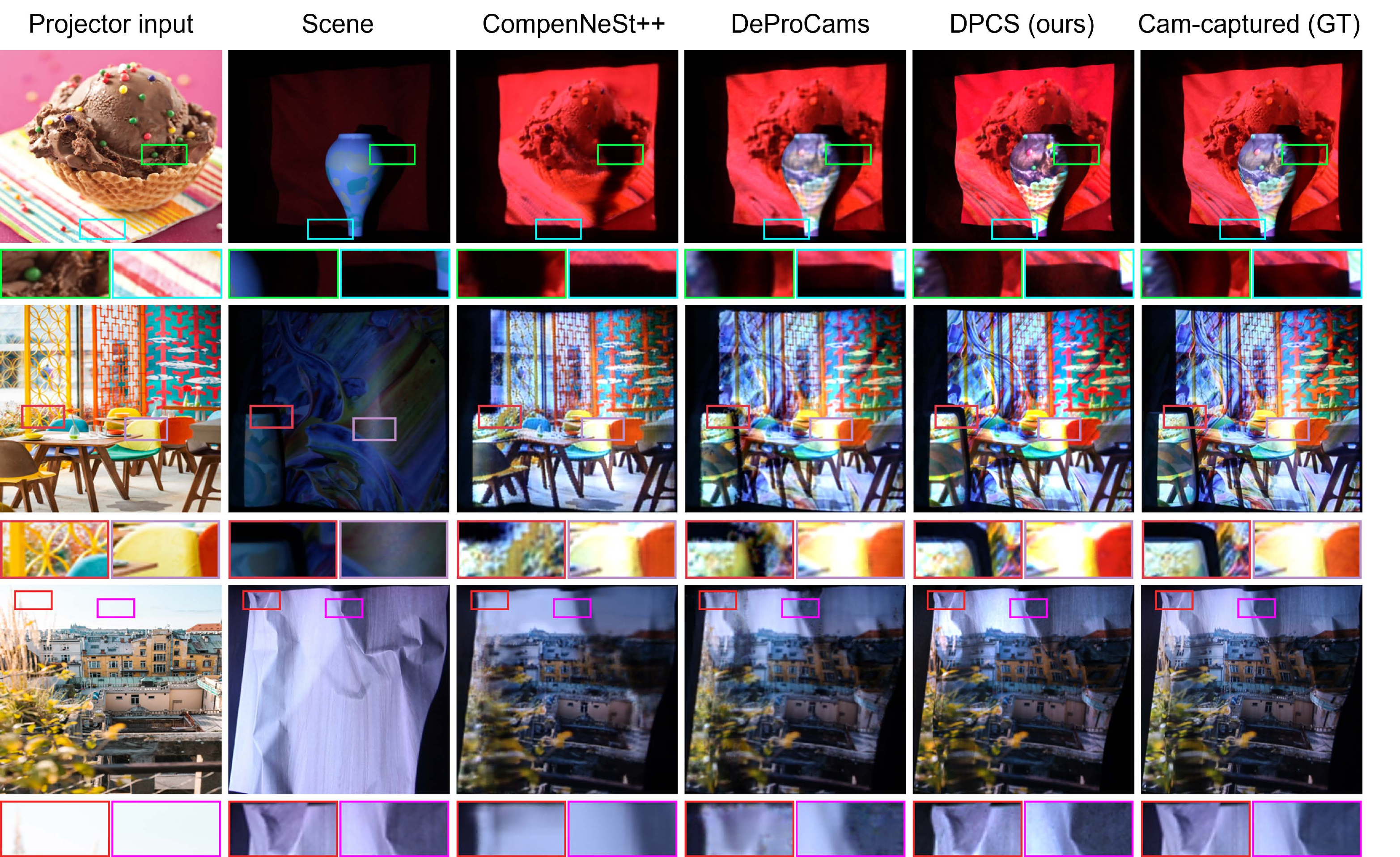}
    \caption{Qualitative comparison on ProCams relighting. We present three scenes under different novel projector input patterns. Each image is provided with two zoomed-in patches for detailed comparison.  The 1\textsuperscript{st} column represents the projector input, the 2\textsuperscript{nd}  shows the camera-captured scenes, the 3\textsuperscript{rd} to 5\textsuperscript{th} present the relighting results of different methods, and the last column is the camera-captured ground truth, i.e., the projection of the 1\textsuperscript{st} onto the 2\textsuperscript{nd}.
}
    \label{fig:relighting}
\end{figure*}

\section{Experiments}
We performed qualitative and quantitative evaluations of our method in different \textbf{real} scenes in ProCams relighting and projector compensation tasks. These scenes vary in surface textures and geometries. The materials in the scene include printed paper with shiny areas and nearly diffuse properties, as well as clothing and other diffuse objects. In our experiments, we used Peak Signal-to-Noise Ratio (PSNR), Structural Similarity Index Measure (SSIM), Learned Perceptual Image Patch Similarity (LPIPS) \cite{zhang2018perceptual} and perceptual color distance \(\Delta\textbf{E}\) \cite{CIE2000} to evaluate the quality of the ProCams relighting and projector compensation.
Our ProCams is calibrated using the software in \cite{huang2020fast} to obtain intrinsics and extrinsics $\Kp, \Kc, \R_\text{p}, \mathbf{t}_\text{p}$. 
Although our DPCS needs only 15 training samples for each scene, for a fair comparison with neural network-based methods, we have captured up to 100 images to compare situations with different numbers of training samples. The scenes were illuminated by an EPSON CB-965 projector with a resolution of 800 $\times$ 600 and captured by a Panasonic LUMIX ZS-220 camera with a resolution of 640 $\times$ 360 in a dark room to ignore the effects of other lighting. We reconstructed the surface point cloud using gray code structured light (SL) and obtained the surface mesh using Poisson surface reconstruction \cite{poisson}. After training, we utilized the optimized scene parameters, as detailed in \cref{fig:overview}, to synthesize camera-captured images under different projections of input patterns. We implemented our method on top of Mitsuba 3’s \texttt{cuda\_ad\_rgb backend} \cite{jakob2022mitsuba3}, with the underlying Dr. Jit \cite{Jakob2020DrJit} framework for forward/reverse-mode AD. 

\subsection{ProCams relighting}\label{sec:simulation result}
For ProCams relighting, we compared our approach with the state-of-the-art method, DeProCams \cite{deprocams}. Additionally, we swapped the input and output of a state-of-the-art projector compensation method, CompenNeSt++ \cite{compennest_pp}, for comparison. All of the methods use the same training and validation data as ours. Note that DPCS requires an additional $42$ SL samples to acquire the geometry, whereas the baseline methods do not. The metrics were calculated for $100$ novel projection patterns and the projection patterns for training and testing were natural images and were not intentionally selected.

The quantitative results are shown in \cref{tab:compare_relit}, and clearly our method outperforms other methods consistently on different numbers of training samples. In particular, even with only $15$ training samples, our method outperforms other methods that use $100$ training samples, because our DPCS leverages SL reconstructed surface and explicit differentiable rendering, while neural networks have to implicitly learn geometry and photometry from a large amount of data. In addition, it implies that $15$ training samples were sufficient for the proposed DPCS to perform the ProCams simulation task well. 

Regarding visual effects in \cref{fig:relighting}, our DPCS significantly outperforms other methods, particularly in shadow handling and indirect lighting, with a notable improvement in regions illuminated by interreflection. CompenNeSt++ \cite{compennest_pp} has the capability to learn some aspects of indirect illumination; however, it faces challenges with complex interreflection, often resulting in smoothed shadow regions and a loss of fine detail. In contrast, DeProCams \cite{deprocams} uses its ShadingNet to learn indirect illumination, requiring an additional image of the surface captured by a camera to encode both direct and indirect reflectance prior. This approach to learning indirect light interactions can be imprecise, as evidenced by distortions in shadowed or illuminated areas outside the projector's FOV. In comparison, DPCS utilizes path tracing-based differentiable rendering, which accurately models multiple reflection paths in a scene, enabling a more realistic simulation of both direct and indirect lighting, particularly in shadowed and interreflection regions, while preserving high-frequency details. Quantitative results presented in \cref{tab:compare_relit} confirm that DPCS considerably surpasses other techniques in LPIPS and SSIM metrics, which can be attributed to its improved rendering of high-frequency details, shadows, and interreflection. Additional comparisons with baseline+SL methods are provided in the supplementary material.

We also compare the computational costs of our method with neural network-based approaches \cite{deprocams,compennest_pp}, as depicted in \cref{tab:compare_time_GPU}, utilizing an Nvidia RTX 3090 GPU. To collect data on resource utilization, we reduced the original CompenNeSt++ batch size to avoid out-of-memory on a single RTX 3090. For other experimental results, the original CompenNeSt++ batch size was used with multiple GPUs. 
Unless otherwise mentioned, we set the SPP of DPCS to $16$ for scene training. Although increasing SPP can improve performance, it also significantly prolongs training time. Therefore, we recommend an SPP of $16$ or higher for balanced results.

\begin{table}[h!]
    \centering
    \caption{Quantitative comparison of real projector compensation. Results are averaged over 10 different setups.}\label{tab:compare compensition}
    \resizebox{\columnwidth}{!}{
    \begin{tabular}{@{}c@{\hskip 5pt}c@{\hskip 5pt}l@{\hskip 5pt}c@{\hskip 5pt}c@{\hskip 5pt}c@{\hskip 5pt}c@{}}
    \toprule
        \textbf{\# Train} & \textbf{\# SL} & \textbf{Model} & \textbf{PSNR}$\uparrow$ & \textbf{SSIM}$\uparrow$ & \textbf{LPIPS}$\downarrow$ & $\Delta$\textbf{E}$\downarrow$ \\
        \midrule
        & 0 & CompenNeSt++  & \textbf{27.4898} &     \textbf{0.8934}     & \textbf{0.0582} &    \textbf{2.1941}   \\
        100 & 0 & DeProCams  & 25.4700 &     0.8728     & 0.0948 &    2.6885   \\
        & 42 & Ours & 26.9341  &    0.8864     & 0.0624 & 2.3334    \\
        \midrule
        & 0 & CompenNeSt++ & \textbf{27.5844} &     \textbf{0.8934} &    \textbf{0.0584}    & \textbf{2.1467}   \\
        50 & 0 & DeProCams & 25.1951 &    0.8715     &0.0966 & 2.7346   \\
        & 42 & Ours & 26.8930 &    0.8842 & 0.0630 &    2.3282    \\
        \midrule
        & 0 & CompenNeSt++ & 26.4883 & \textbf{0.8912} &    0.0681 &    \textbf{2.3088}  \\
        15 & 0 & DeProCams & 21.6648 &    0.8524 &    0.1211 & 3.4482   \\
        & 42 & Ours & \textbf{27.0638} &    0.8855 &    \textbf{0.0629}     & 2.3321     \\
        \midrule
        & 0 & CompenNeSt++ & 25.5781 &    \textbf{0.8851} &    0.0794 &    2.5066    \\
        5 & 0 & DeProCams & 14.5349 &     0.7753     & 0.2385 & 7.7353   \\
        & 42 & Ours &\textbf{26.8080}  &     0.8811 &    \textbf{0.0681}     &\textbf{2.4501}    \\
    \bottomrule
    \end{tabular}}
\end{table}

\begin{figure*}[h!]
    \centering
    \includegraphics[width=0.99\textwidth]{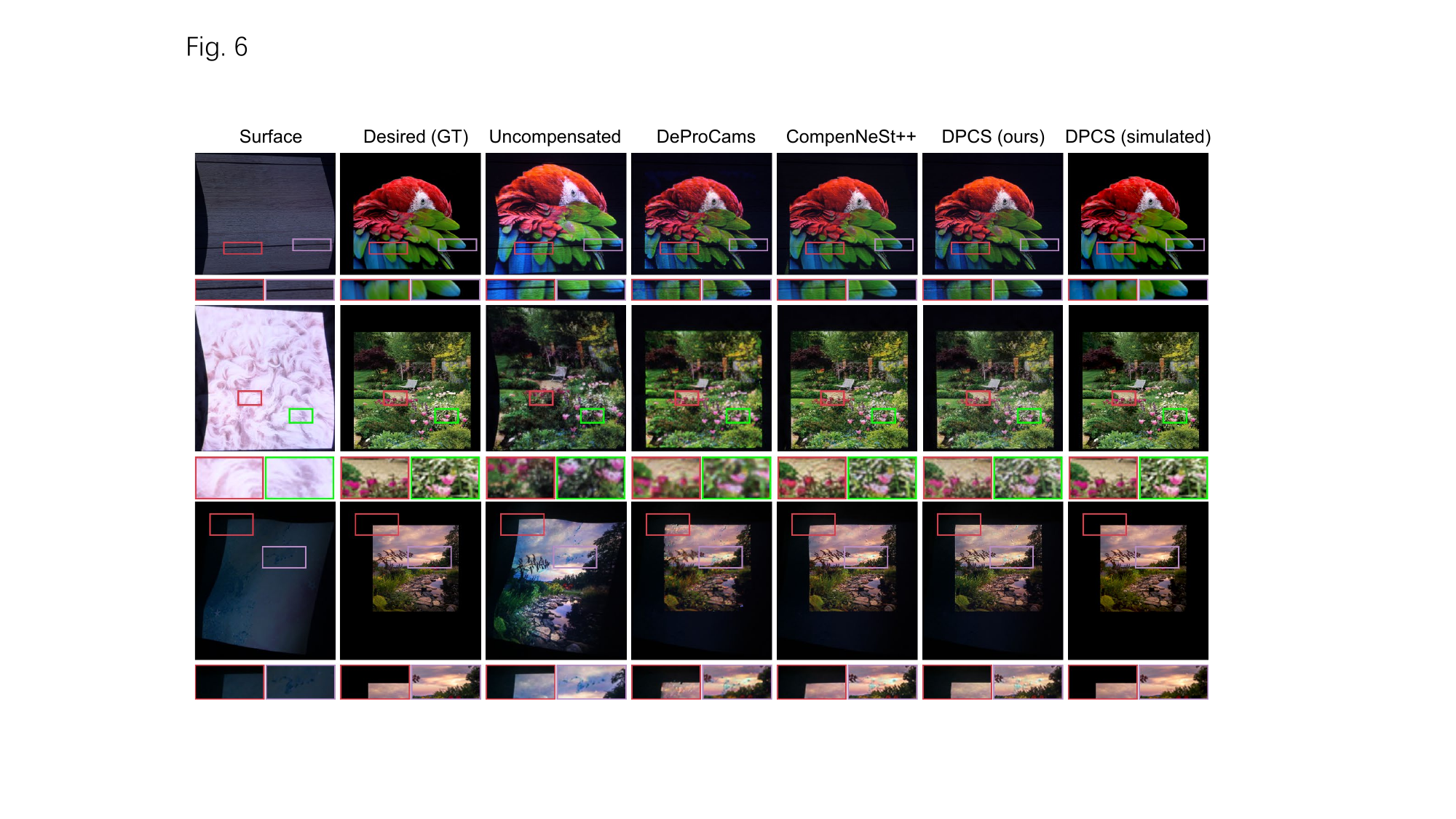}
    \caption{Qualitative comparison of real projector compensation. Columns $1$ to $3$ display the projection surface, the desired image as perceived by viewers, and the uncompensated projection, respectively. Subsequent columns present real camera-captured outcomes from various compensation techniques. DPCS (simulated) represents the simulated compensated result in the renderer.}
    \label{fig:compen_result}
\end{figure*}

\begin{figure*}[!h]
    \centering
    \includegraphics[width=1\textwidth]{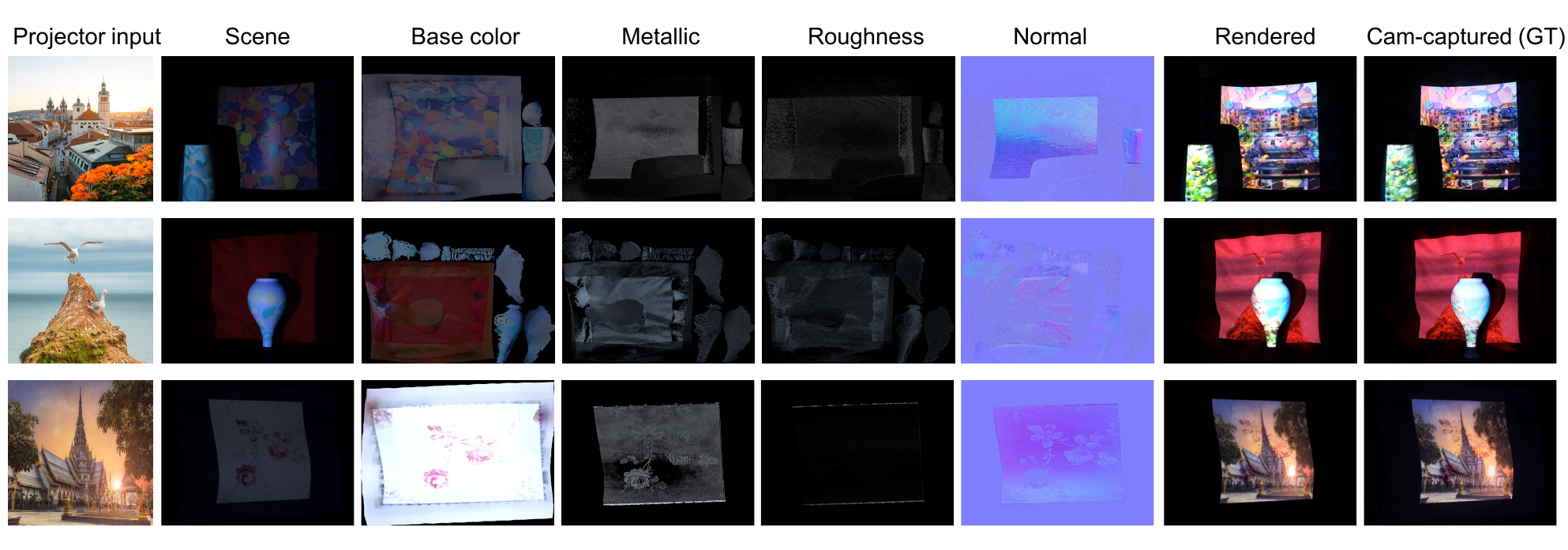}
    \caption{BRDF estimation and relighting of DPCS for different scenes. Our method decomposes the surface material into interpretable quantities which are passed into a Disney principled \cite{Bur12} to generate the final image. Note that BRDF maps are flattened into 2D texture space using UV unwrapping, and the values of the normal map are specified relative to the surface normal, following Mitsuba. For example, a value of $(0, 0, 1)$ in the normal map signifies no change to the surface normal. }
    \label{fig:decomposition}
\end{figure*}

\begin{figure*}[t]
  \centering
  \includegraphics[width=1\textwidth]{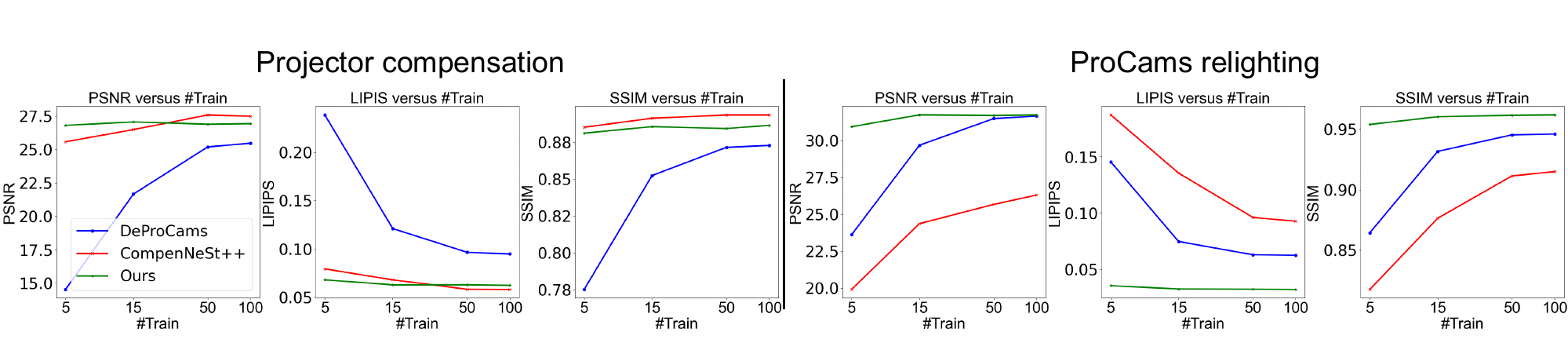}
  \caption{Performance of ProCams relighting and projector compensation under different number of training images (\#~Train) in \cref{tab:compare_relit} and \cref{tab:compare compensition}, respectively.}
  \label{fig:metric_compare}
\end{figure*}

\subsection{Projector compensation}
\label{sec: compensation_result}
Following the optimization of the scene parameters, we achieved a successful simulation of the ProCams scene within DPCS. This virtual representation allows us to fine-tune specific physical parameters for distinct SAR applications. By minimizing the pixel loss between DPCS rendered images and those expected from projection and captures of the real scene, we can obtain a compensated projector input image using gradient descent. The implementation of this process is illustrated in \cref{fig:compensation_pipeline}.
We compare our method against two state-of-the-art neural network-based projector compensation methods DeProCams \cite{deprocams} and CompenNeSt++ \cite{compennest_pp} in a real-world projector compensation task. Each method generates compensated projector inputs by projecting them and then capturing them with a camera. The captured compensation results are compared against the target ground truth. Qualitative comparisons are illustrated in \cref{fig:compen_result}. We can observe that our DPCS performs similarly to CompenNeSt++. However, on some very difficult-to-compensate dark textures, our method achieves good results where CompenNeSt++ and DeProCams may fail at these regions. The quantitative results are shown in \cref{tab:compare compensition}. Note that our method outperforms DeProCams when using a different number of training samples, and is very close to CompenNeSt++. Furthermore, with a smaller number of training samples, e.g., \#Train$\leq 15$, our DPCS has much smaller pixel errors (PSNR/RMSE) compared to CompenNeSt++. Unlike our DPCS, CompenNeSt++ is specifically designed for projector compensation, and cannot perform other SAR tasks, such as BRDF, CRF/PRF estimation, and novel scene simulation. Interestingly, although our DPCS excels in acquiring high-frequency details, it presents some color inaccuracies in low-frequency areas compared to neural network methods. This issue may stem from the unmodeled chromatic aberration of the projector and the limited real-world applicability of the BRDF model, which deep learning can implicitly correct using its network parameters. As a result, explicit modeling of chromatic aberration in projections and developing a more accurate BRDF model for ProCams may be a promising direction for future research.

We conduct a more detailed examination of DPCS performance across a varying number of training samples. The results displayed in \cref{fig:metric_compare} demonstrate that DPCS can be adequately trained with just $15$ training samples. This outcome stands in favorable comparison to various neural network-based approaches in two subsequent tasks: ProCams relighting and projector compensation. It is important to note that the $x$ axes are presented on a logarithmic scale for improved clarity.

\subsection{BRDF estimation}
After training, our DPCS can estimate the BRDF maps of the scene $\mathbf{m}$, which include roughness $\in [0, 1]$, metallic $\in [0, 1]$, base color $\in [0, 1]^3$, and an extra normal map $\Vn$ $\in [0, 1]^3$ for finer adjustments. The results in \cref{fig:decomposition} show that the projection patterns are not baked into the BRDF maps, and the rendered images look close to the GT captured by the camera. Moreover, it is noticeable in the BRDF map that, aside from the projector directly illuminated regions where the BRDF is sampled relatively well, the indirect light regions have a sparser distribution of sampling rays. The pixel values around these regions may be smoothed due to the smoothing loss (\cref{eq:optimization}). Despite these results, our approach still faces challenges in accurately estimating real-world ProCams BRDF maps. As shown in \cref{fig:decomposition}, interdependencies frequently exist between different BRDF maps estimated by DPCS; for example, the base color may be coupled with other BRDF maps. Incorporating additional data, such as shadows or specular highlights from multi-view projections, to constrain the optimization of BRDF maps during differentiable rendering may help mitigate this problem.

\subsection{Novel scene simulation}
\begin{figure}[t]
  \centering
  \includegraphics[width=1.0\columnwidth]{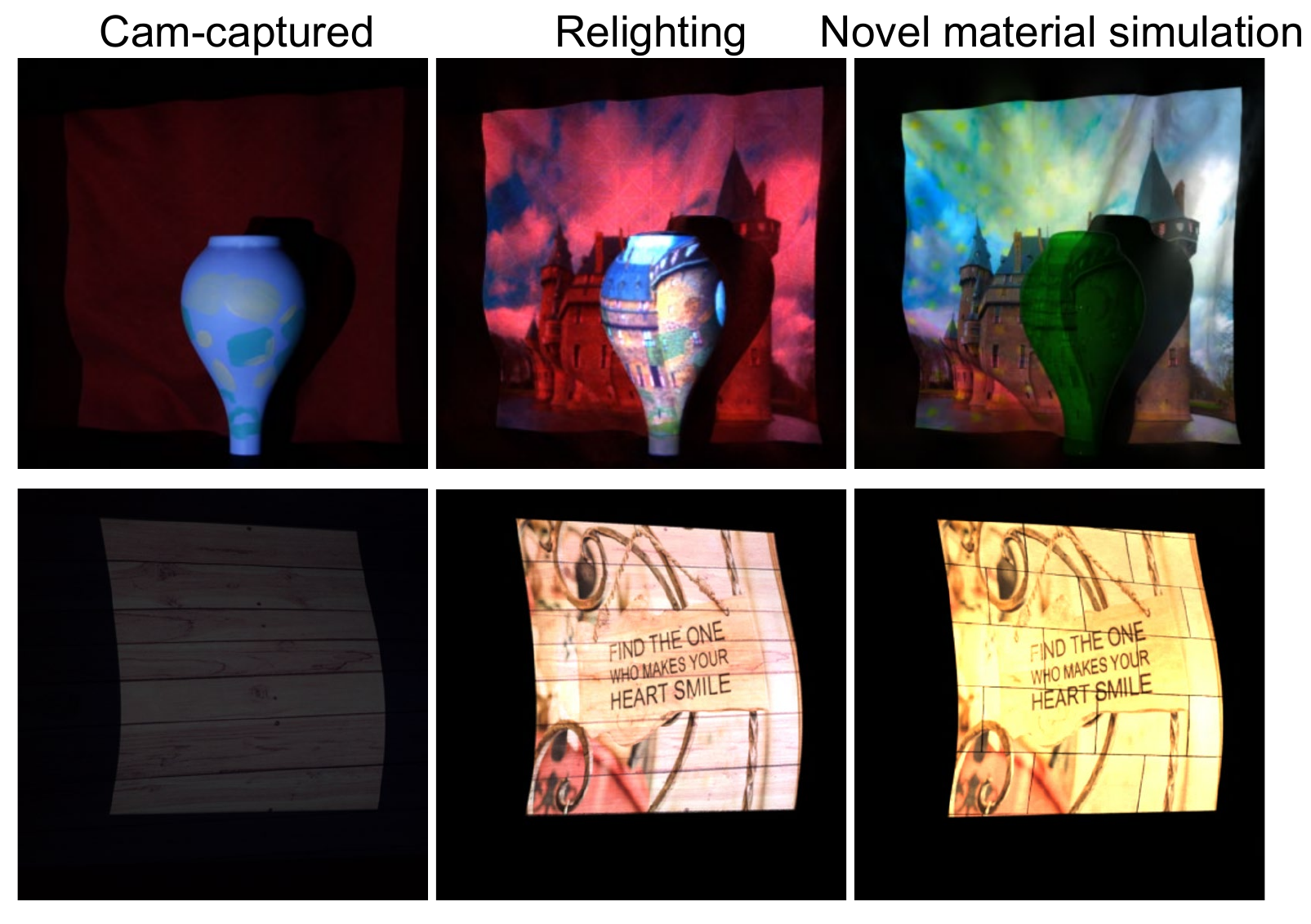}
  \caption{Novel material simulation. We present two scenes in two rows, modifying the surface's BRDF to a different PBR material and projecting the same pattern to it.}
  \label{fig:novel appearance}
\end{figure}

\begin{figure}[t]
    \centering
    \includegraphics[width=1\columnwidth]{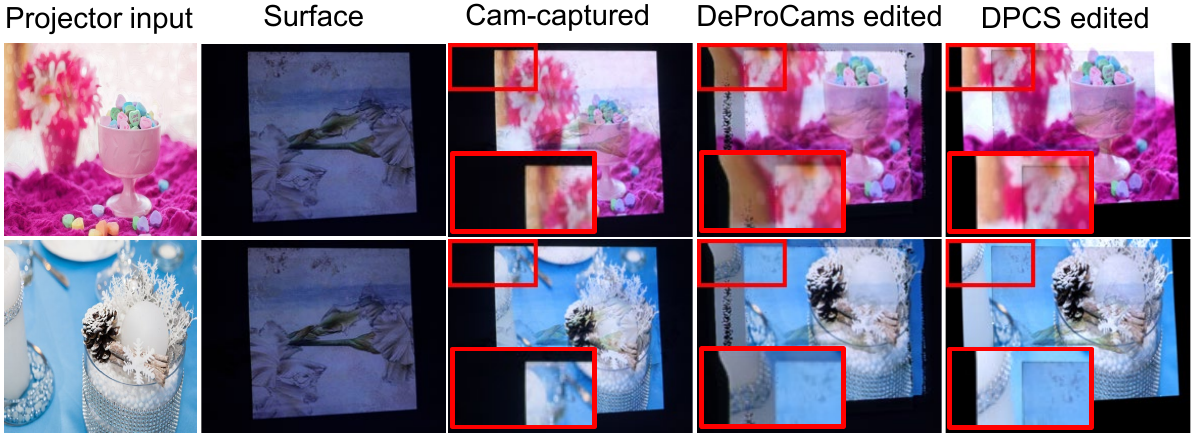}
    \caption{Novel projector pose and FOV synthesize. We edit both the projector intrinsics and extrinsics. The 2\textsuperscript{nd} column represents the real camera-captured surface. The 3\textsuperscript{rd} column is the real camera-captured scene (at the original projector pose) under the projection in the 1\textsuperscript{st} column. The last two columns are the synthesized results when we move the projector up and increase its FOV.}
    \label{fig:scene_edit}
\end{figure}

Novel scene simulation aims to virtually modify scene parameters, such as surface BRDF, ProCams response functions, poses, and focal lengths. This technique is crucial for projection mapping, as simulating the visual effects before deployment to uncalibrated scenes can greatly reduce human effort. For example, DPCS enables the user to synthesize new projectors with different intrinsics and extrinsics. We can virtually move the projector up and increase the projector's field of view (FOV). Qualitative comparisons of this novel scene simulation with DeProCams are shown in \cref{fig:scene_edit}. It is evident that the DeProCams synthesized novel scene has distortions at the edges of the projector FOV. By contrast, our DPCS-synthesized scenes are more realistic. This is because our method is physically-based, and the extrinsics only affect the virtual projector pose. 
The novel material simulation is shown in \cref{fig:novel appearance}, where we virtually change the surface BRDF. Note that DeProCams \cite{deprocams} cannot perform this task, since it only estimates \textit{rough} shading material and a depth map. 
By contrast, our DPCS represents object geometries using standard triangle meshes, allowing explicit geometry editing. For example, the geometry in a virtual scene can be modified to simulate projection effects on a moving surface, such as a sheet billowing in the wind. In addition, this approach can simulate compensation for changes in geometry. Please refer to the supplementary video, which demonstrates our DPCS applied to compensate for a morphing surface.

\begin{figure*}[!th]
    \centering
    \includegraphics[width=1\textwidth]{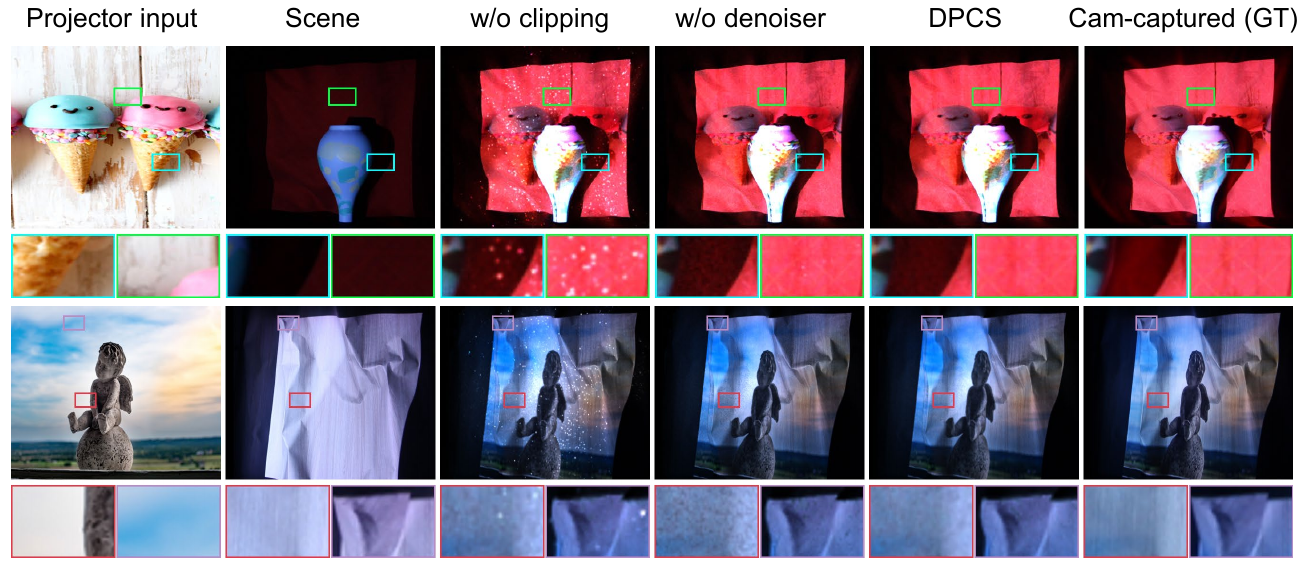}
    \caption{Ablation study of DPCS. The 1\textsuperscript{st} column represents the input to the projector, the 2\textsuperscript{nd} column shows the camera-captured scenes, the 3\textsuperscript{rd} and 4\textsuperscript{th} columns respectively present the relighting results of DPCS without radiance clipping operation and without denoising filter, the 5\textsuperscript{th} is the DPCS relighting result and the last column is the camera-captured ground truth. }
    \label{fig:ablation study}
\end{figure*}

\subsection{Ablation study}
We performed ablation studies on the cross-bilateral denoising filter and the radiance clipping operation mentioned in \cref{sec:radiance clipping}. Specifically, we remove the denoising filter and the radiance clipping from our framework and name them \textit{w/o denoiser} and \textit{w/o clipping}, respectively. The quantitative results of the relighting are presented in \cref{tab:compare_relit}, while the qualitative comparisons are illustrated in \cref{fig:ablation study}. We found that DPCS and DPCS \textit{w/o denoiser} performed similarly in quantitative results (less than $1\%$ difference).
However, employing such a denoiser leads to a reduced local variance in qualitative results, especially for the area with indirect illumination for its uneven sampling (see \autoref{fig:ablation study}), which looks more visually pleasing. We use it in the framework to reduce local variance in the image, aligning better with the viewer's visual experience. However, simply using a filter for denoising does not improve the overall image quality, as it primarily focuses on local variance details within a window size. Training a specialized denoising network \cite{OpenImageDenoise} to learn the noise distribution of Monte Carlo sampling in ProCams scenes may further enhance this aspect. 

For \textit{w/o clipping}, low-probability light path samples may contribute significantly to the brightness of certain pixels during Monte Carlo integration, leading to ``firefly” noise. By applying a radiance clipping operation, the radiance captured by the camera after (inter)reflections is constrained within the radiance range initially emitted by the projector, effectively suppressing this kind of noise.

\section{Limitations and Future Work}
Our method does not utilize multiple viewpoints of images as references, while directly optimizing the mesh in a single-view setting may lead to various ambiguities, causing the model to become trapped in local optima. 
Therefore, we use Gray-code structured light to capture the projection surface geometry; this may be more cumbersome than previous end-to-end methods \cite{deprocams}. 
Future work may involve guiding mesh optimization using multi-view perspectives or leveraging additional information, such as shadows and interreflection or refraction \cite{Reparams,nicolet2021large} cast by the projector’s light on the scene, would make the simulation of the projection scene more accurate and meaningful.
Additionally, although we use path tracing-based differentiable rendering to simulate complex projection light interaction effects, the precise decomposition of different BRDF maps remains an under-constrained problem. For example, specular highlights can get "baked" on different maps. We plan to further decompose and obtain more accurate BRDFs by introducing constraints from additional information such as shadows and highlights generated from multi-view projections. Furthermore, although DPCS excels in simulating complex projection effects, handling fewer training samples, and supporting scene parameters editing, its high computational demands significantly limit its suitability for real-time applications. This study does not achieve real-time performance, but a future direction could be optimizing Monte Carlo denoising algorithms to reduce the computational overhead introduced by the denoising process itself. By achieving this optimization, a real-time denoising algorithm allows for a reduced sample-per-pixel (spp) count, balancing the trade-off between rendering quality and real-time performance.

\section{Conclusion}

We propose DPCS, a path tracing-based differentiable simulation method for ProCams. DPCS can solve for nonlinear responses, exposure/gain, and material in the ProCams setup. This allows us to efficiently modify certain parameters of the systems to perform novel scene simulations, which can be used to synthesize more realistic projection mapping effects in virtual setups for SAR applications. With only $15$ training samples, DPCS can simulate and test complex projection mapping scenes with interreflection to optimize the projection mapping effect. Meanwhile, data synthesized with this system enables more precise and efficient data acquisition of ProCams through a complex light transport approach. It would be interesting to explore how to optimize the mesh under multi-view constraints or by leveraging additional information, such as shadows cast by the projector light on the scene. This integration could lead to more realistic simulations and facilitate downstream tasks from novel perspectives.

\acknowledgments{
We thank the anonymous reviewers for valuable and inspiring comments and suggestions. 
}

\bibliographystyle{abbrv-doi-hyperref}

\bibliography{dpcs_arxiv}

\clearpage
\maketitlesupplementary
\appendix  
\setcounter{page}{1}
\section{Introduction}
\begin{figure*}[h!]
    \centering
    \includegraphics[width=1.01\textwidth]{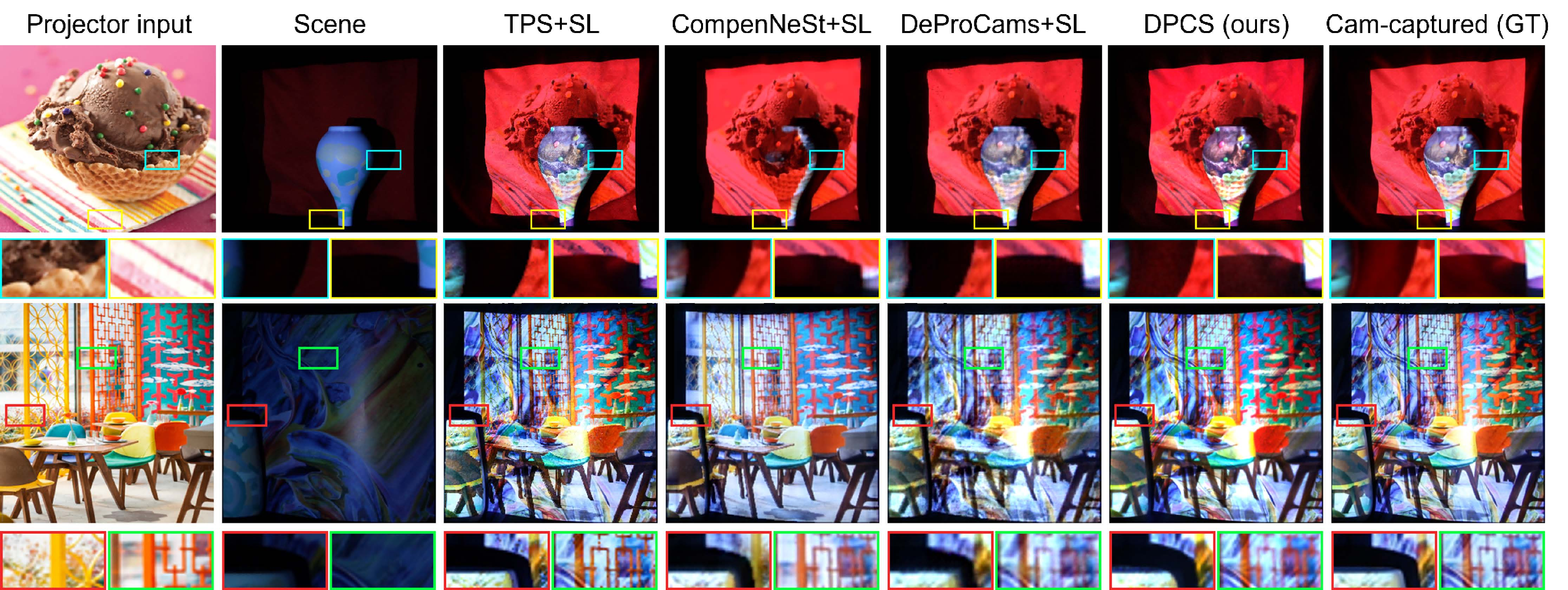}
    \caption{Qualitative comparison of our method with baseline+SL methods on \textbf{ProCams relighting}. We present two scenes under different novel projector input patterns. Each image is provided with two zoomed-in patches for detailed comparison.  The $1^{st}$ column represents the projector input, the $2^{nd}$  shows the camera-captured scenes, the $3^{rd}$ to $6^{th}$ present the relighting results of different methods, and the last column is the camera-captured ground truth, i.e., the projection of the $1^{st}$ onto the $2^{nd}$.
}
    \label{fig:sup_relighting_sl}
\end{figure*}

\begin{figure*}[h!]
    \centering
    \includegraphics[width=1.01\textwidth]{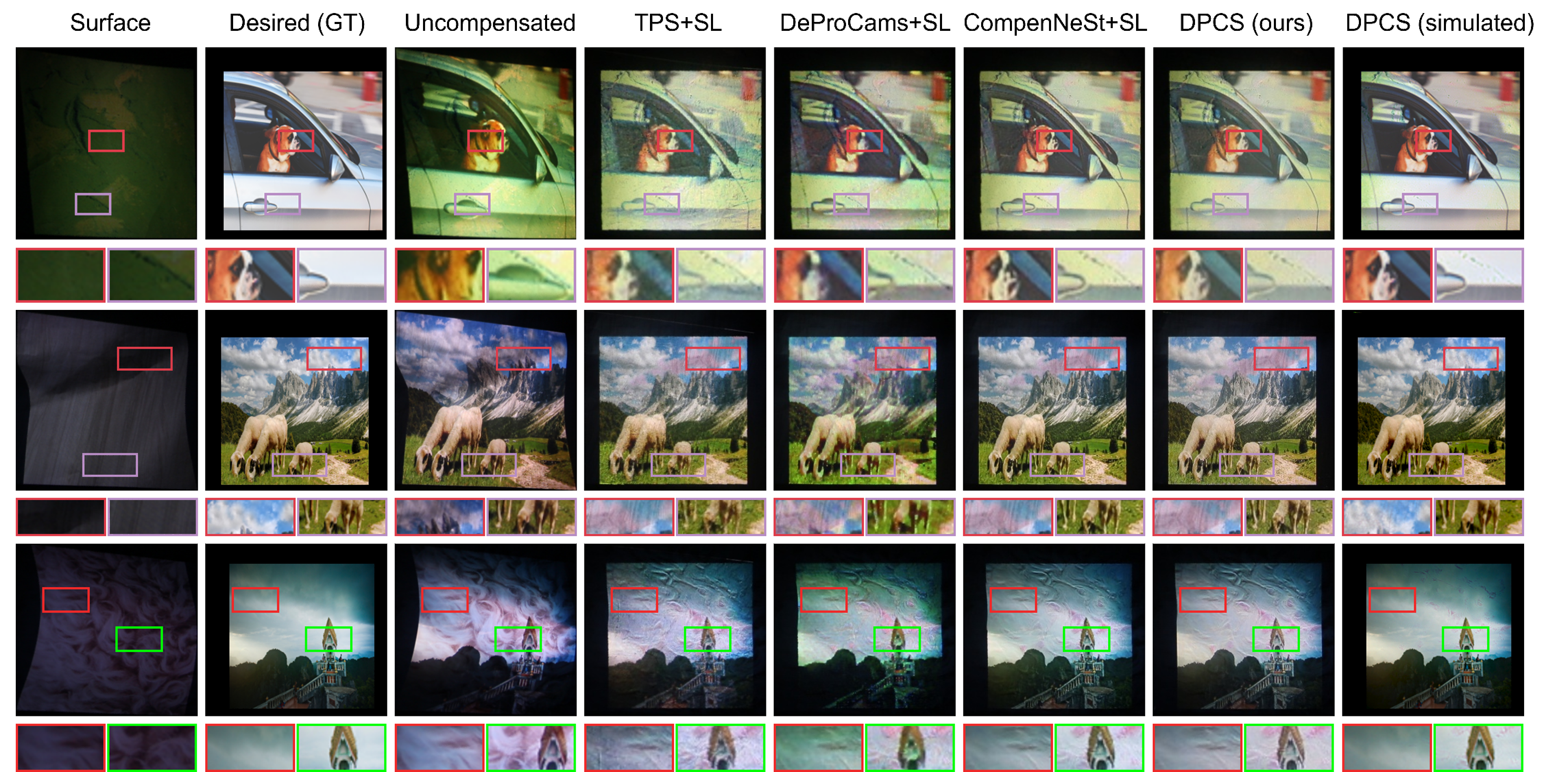}
    \caption{Qualitative comparison of our method with baseline+SL methods on real \textbf{projector compensation}. Columns $1$ to $3$ display the projection surface, the desired image as perceived by viewers, and the uncompensated projection, respectively. Subsequent columns present real camera-captured outcomes from various compensation techniques. DPCS (simulated) represents the simulated compensated result in renderer.
}
    \label{fig:sup_compensation}
\end{figure*}

\maketitle
In this supplementary material, we compared the methods (baseline+SL) on \textbf{ProCams relighting} and \textbf{projector compensation}. Specifically, CompenNeSt+SL uses structured light (SL) rectification combined with CompenNeSt; DeProCams + SL uses SL depth instead of learning it from sample images and freezes depth during DeProCams training; TPS+SL incorporates thin plate splines \cite{tps} with SL rectification. For \textbf{ProCams relighting}, we made comparisons between the 14 datasets summarized in Table 2 of the main paper. The quantitative comparison is shown in \autoref{tab:sup_compare_relit}, and the qualitative comparisons on \textbf{ProCams relighting} with baseline+SL methods are demonstrated in \autoref{fig:sup_relighting_sl}. For \textbf{projector compensation}, as this task cannot be reproduced using the same data as in the main paper due to real setup changes, we prepared 5 new setups for comparison, and the quantitative and qualitative comparisons are shown in \autoref{tab:sup_compare compensition} and \autoref{fig:sup_compensation}, respectively.

\begin{figure*}[h!]
    \centering
    \includegraphics[width=0.99\textwidth]{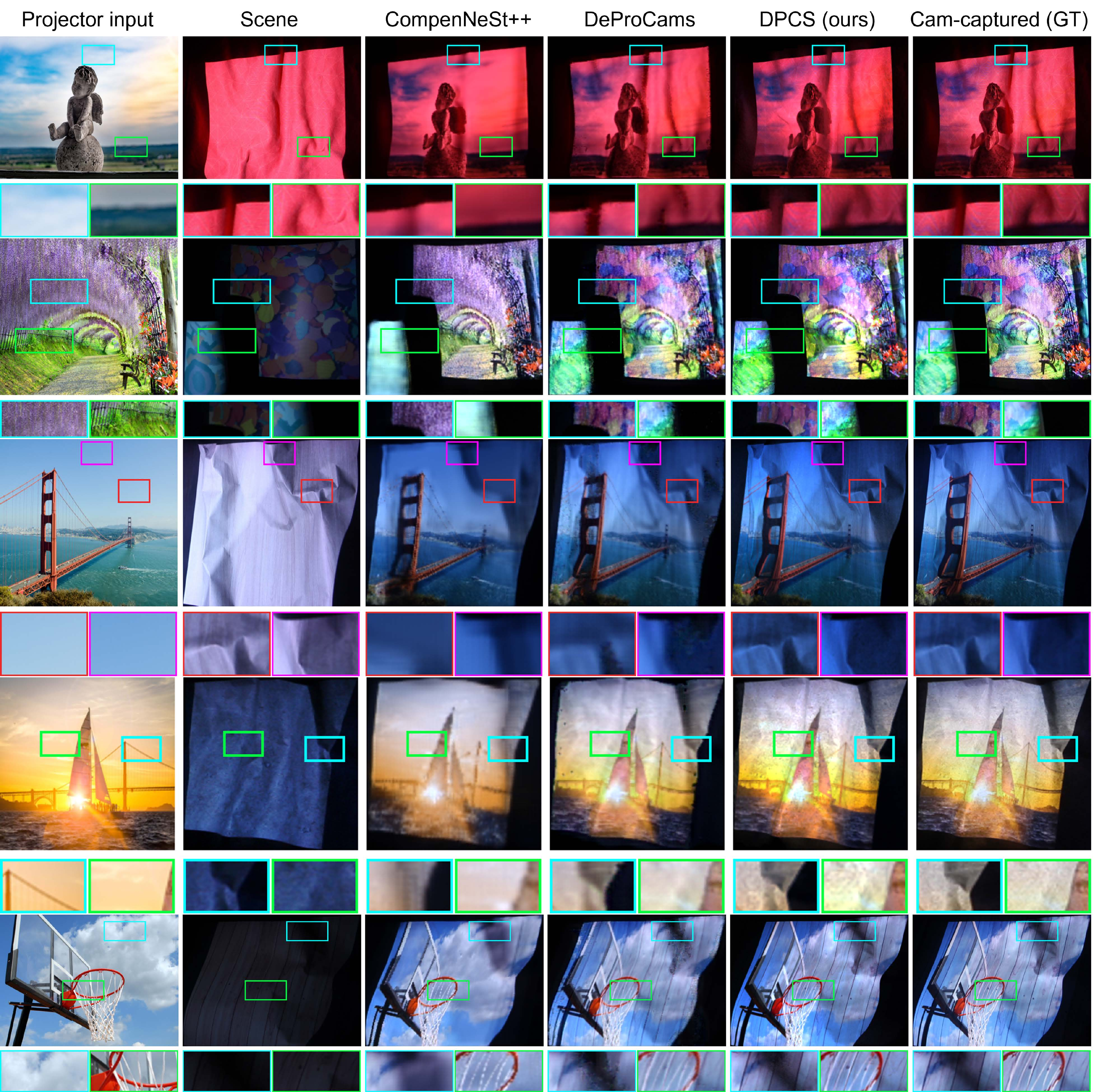}
    \caption{Qualitative comparison on \textbf{ProCams relighting}. We present several scenes under different novel projector input patterns. Each image is provided with two zoomed-in patches for detailed comparison.  The $1^{st}$ column represents the projector input, the $2^{nd}$  shows the camera-captured scenes, the $3^{rd}$ to $5^{th}$ present the relighting results of different methods, and the last column is the camera-captured ground truth, i.e., the projection of the $1^{st}$ onto the $2^{nd}$.
}
    \label{fig:sup_relighting_1}
\end{figure*}

\begin{figure*}[h!]
    \centering
    \includegraphics[width=0.99\textwidth]{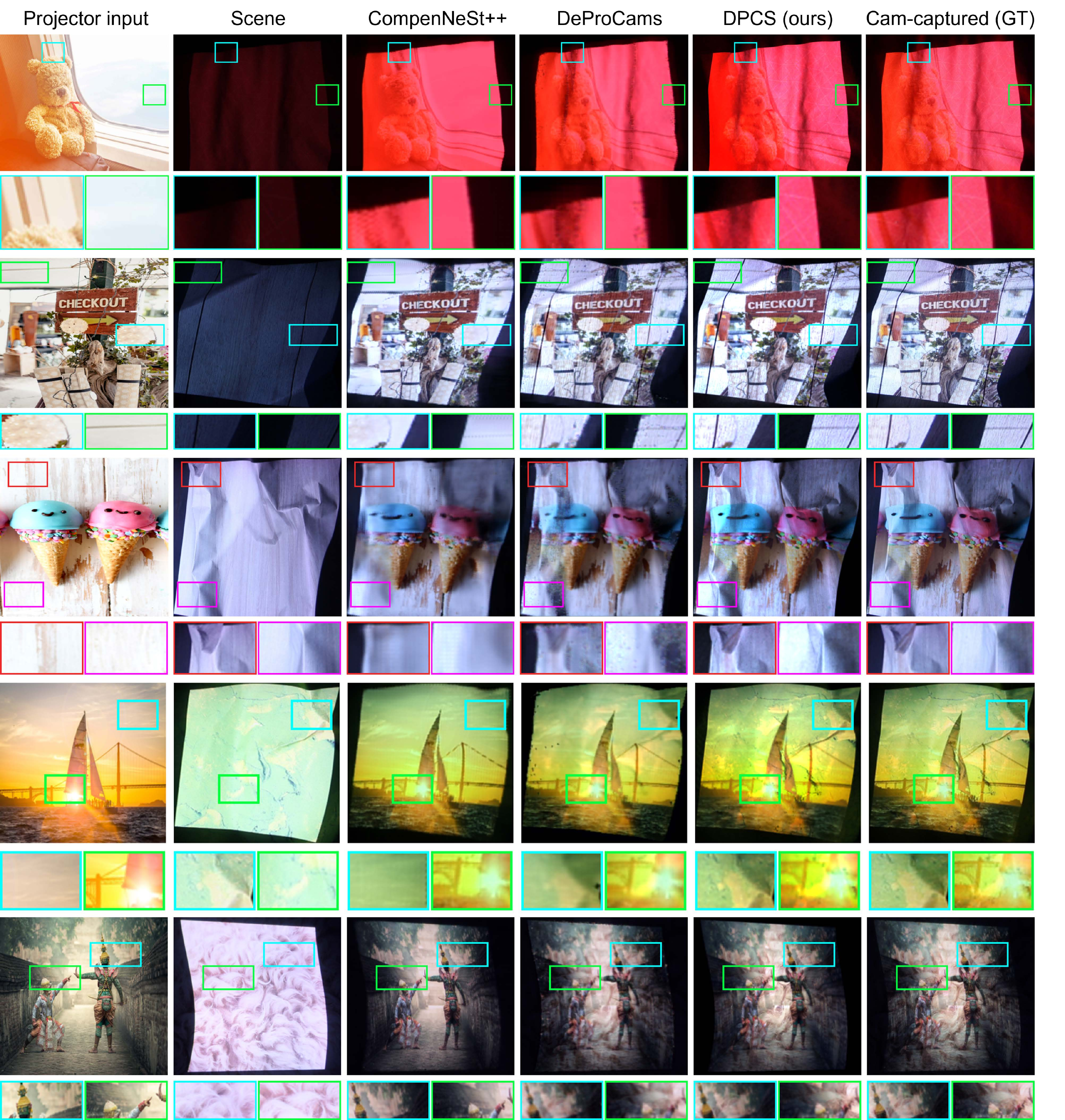}
    \caption{Qualitative comparison on \textbf{ProCams relighting}. We present several scenes under different novel projector input patterns. Each image is provided with two zoomed-in patches for detailed comparison.  The $1^{st}$ column represents the projector input, the $2^{nd}$  shows the camera-captured scenes, the $3^{rd}$ to $5^{th}$ present the relighting results of different methods, and the last column is the camera-captured ground truth, i.e., the projection of the $1^{st}$ onto the $2^{nd}$ .}
    \label{fig:sup_relighting_2}
\end{figure*}

\begin{figure*}[h!]
    \centering
    \includegraphics[width=0.99\textwidth]{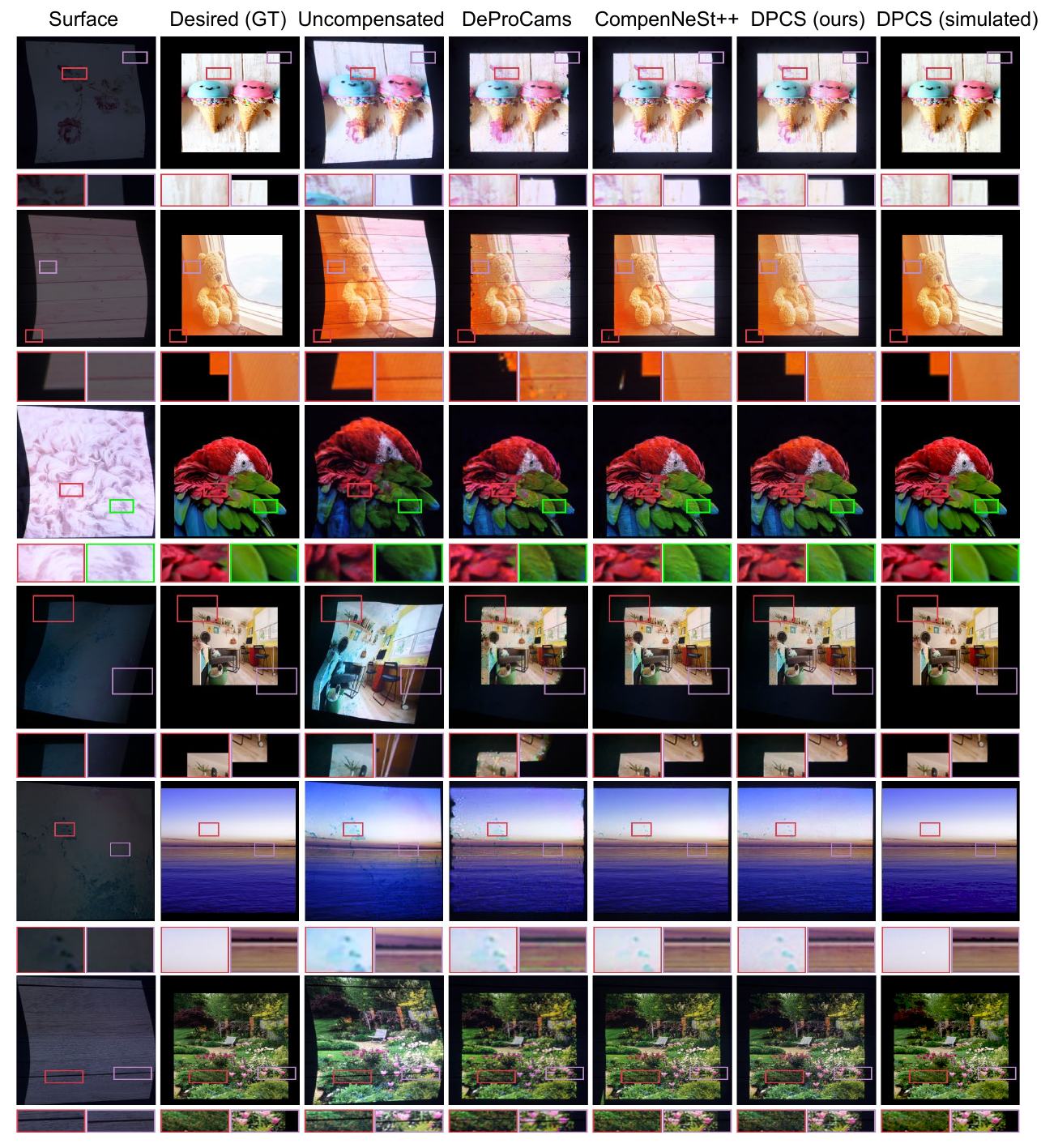}
    \caption{Qualitative comparison of real \textbf{projector compensation}. Columns $1$ to $3$ display the projection surface, the desired image as perceived by viewers, and the uncompensated projection, respectively. Subsequent columns present real camera-captured outcomes from various compensation techniques. DPCS (simulated) represents the simulated compensated result in renderer.}
    \label{fig:sup_compen_result}
\end{figure*}

We also showed more experimental results: qualitative comparison with the different ProCams simulation method \cite{deprocams,compennest_pp} on \textbf{ProCams relighting} in \autoref{fig:sup_relighting_1} and \autoref{fig:sup_relighting_2}, qualitative comparison with the learning-based state-of-the-art methods \cite{deprocams,compennest_pp} on real camera-captured compensation in \autoref{fig:sup_compen_result}.
\FloatBarrier
\begin{table}[!h]
    \centering
    \caption{Quantitative comparison of our method with baseline+SL methods on real \textbf{projector compensation}. Results are averaged over 5 different setups.}
    \label{tab:sup_compare compensition}
    \resizebox{\columnwidth}{!}{
    \begin{tabular}{c@{\hskip 5pt}l@{\hskip 5pt}c@{\hskip 5pt}c@{\hskip 5pt}c@{\hskip 5pt}c}
    \toprule
    \textbf{\# Train} & \textbf{Model} & \textbf{PSNR}$\uparrow$ & \textbf{SSIM}$\uparrow$ & \textbf{LPIPS}$\downarrow$ & $\boldsymbol{\Delta}$\textbf{E}$\downarrow$ \\
    \midrule
    100 
    & CompenNeSt++       & \textbf{28.3088} & \textbf{0.9173} & 0.0497 & \textbf{1.8488} \\
    & CompenNeSt+SL     & 28.1167 & 0.9153 & \textbf{0.0488} & 1.9555 \\
    & DeProCams          & 26.7240 & 0.9091 & 0.0727 & 2.1270 \\
    & DeProCams+SL        & 25.7595 & 0.8973 & 0.0795 & 2.3777 \\
    & TPS+SL             & 26.7408 & 0.8851 & 0.0786 & 2.1085 \\
    & DPCS (ours)        & 27.4364 & 0.9101 & 0.0559 & 1.9609 \\
    \midrule
    50
    & CompenNeSt++       & \textbf{28.2789} & \textbf{0.9175} & 0.0502 & \textbf{1.8824} \\
    & CompenNeSt+SL     & 28.1649 & 0.9158 & \textbf{0.0488} & 1.9110 \\
    & DeProCams          & 26.7472 & 0.9092 & 0.0719 & 2.0545 \\
    & DeProCams+SL        & 25.8343 & 0.8981 & 0.0791 & 2.4384 \\
    & TPS+SL             & 26.4994 & 0.8808 & 0.0827 & 2.1296 \\
    & DPCS (ours)        & 27.4112 & 0.9094 & 0.0568 & 1.9835 \\
    \midrule
    15
    & CompenNeSt++       & 27.6155 & 0.9126 & 0.0640 & 1.9899 \\
    & CompenNeSt+SL     & \textbf{28.0485} & \textbf{0.9144} & \textbf{0.0505} & \textbf{1.9502} \\
    & DeProCams          & 25.6273 & 0.9037 & 0.0803 & 2.2375 \\
    & DeProCams+SL        & 25.8428 & 0.8976 & 0.0790 & 2.3682 \\
    & TPS+SL             & 25.8990 & 0.8687 & 0.0970 & 2.2568 \\
    & DPCS (ours)        & 27.4555 & 0.9082 & 0.0576 & 1.9955 \\
    \midrule
    5
    & CompenNeSt++       & 26.6723 & 0.9063 & 0.0729 & 2.2798 \\
    & CompenNeSt+SL     & \textbf{27.5485} & \textbf{0.9095} & \textbf{0.0573} & \textbf{2.0965} \\
    & DeProCams          & 18.5602 & 0.8503 & 0.1460 & 4.4061 \\
    & DeProCams+SL        & 24.9926 & 0.8948 & 0.0847 & 2.7301 \\
    & TPS+SL             & 21.4206 & 0.8310 & 0.1711 & 3.5088 \\
    & DPCS (ours)        & 26.9283 & 0.9024 & 0.0642 & 2.1874 \\
    \bottomrule
    \end{tabular}}
\end{table}
\FloatBarrier
\begin{table}[!t]
    \vspace{-16.5\baselineskip}
    \centering
    \caption{Quantitative comparison of our method with baseline+SL methods on \textbf{ProCams relighting}. Results are averaged over 14 different setups in the main paper Table 2. Note that the results are very slightly different from the original main paper Table 2 due to random seeds, but they do not affect this paper's conclusion.}
    \label{tab:sup_compare_relit}
    \resizebox{\columnwidth}{!}{
    \begin{tabular}{c@{\hskip 5pt}l@{\hskip 5pt}c@{\hskip 5pt}c@{\hskip 5pt}c@{\hskip 5pt}c}
    \toprule
    \textbf{\# Train} & \textbf{Model} & \textbf{PSNR}$\uparrow$ & \textbf{SSIM}$\uparrow$ & \textbf{LPIPS}$\downarrow$ & $\boldsymbol{\Delta}$\textbf{E}$\downarrow$\\
    \midrule
    100 
    & CompenNeSt++        & 26.1202    & 0.9127    & 0.0975    & 2.2828 \\
    & CompenNeSt+SL         & 27.2156 & 0.9304 & 0.0680 & 2.0698 \\
    & DeProCams           & 31.7081    & 0.9456    & 0.0612    & 1.3095 \\
    & DeProCams+SL         & 31.4722    & 0.9497    & 0.0443    & 1.4249 \\
    & TPS+SL                & 29.4055 & 0.9289 & 0.0708 & 1.4793 \\
    & DPCS (ours)         & 31.8080    & 0.9628    & 0.0318    & \textbf{1.2607} \\
    & w/o denoiser (ours) & \textbf{31.8112} & \textbf{0.9635} & \textbf{0.0238} & 1.2826 \\
    & w/o clipping (ours) & 30.9332    & 0.9527    & 0.0510    & 1.3199 \\
    \midrule
    50 
    & CompenNeSt++        & 26.7989    & 0.9170    & 0.0917    & 2.1466 \\
    & CompenNeSt+SL         & 26.9667 & 0.9265 & 0.0688 & 2.1289 \\
    & DeProCams           & 31.6300    & 0.9465    & 0.0608    & 1.2960 \\
    & DeProCams+SL         & \textbf{32.3082} & 0.9548    & 0.0420    & 1.3358 \\
    & TPS+SL                & 29.0233 & 0.9228 & 0.0765 & 1.5784 \\
    & DPCS (ours)         & 31.8616    & 0.9625    & 0.0317    & \textbf{1.2611} \\
    & w/o denoiser (ours) & 31.8544    & \textbf{0.9634} & \textbf{0.0238} & 1.2835 \\
    & w/o clipping (ours) & 31.0299    & 0.9529    & 0.0502    & 1.3189 \\
    \midrule
    15 
    & CompenNeSt++        & 24.5035    & 0.8908    & 0.1194    & 2.8091 \\
    & CompenNeSt+SL         & 25.7810 & 0.9135 & 0.0734 & 2.4687 \\
    & DeProCams           & 29.7267    & 0.9319    & 0.0736    & 1.5762 \\
    & DeProCams+SL         & \textbf{31.9049} & 0.9512    & 0.0438    & 1.4024 \\
    & TPS+SL                & 27.5325 & 0.8970 & 0.1031 & 2.0077 \\
    & DPCS (ours)         & 31.8618    & 0.9612    & 0.0321    & \textbf{1.3202} \\
    & w/o denoiser (ours) & 31.8623    & \textbf{0.9618} & \textbf{0.0250} & 1.3370 \\
    & w/o clipping (ours) & 30.9345    & 0.9509    & 0.0517    & 1.3836 \\
    \midrule
    5 
    & CompenNeSt++        & 21.7428    & 0.8401    & 0.1647    & 3.8288 \\
    & CompenNeSt+SL         & 24.1217 & 0.8943 & 0.0858 & 3.0168 \\
    & DeProCams           & 23.7797    & 0.8661    & 0.1433    & 3.2590 \\
    & DeProCams+SL         & 30.2657    & 0.9424    & 0.0514    & 1.7480 \\
    & TPS+SL                & 20.4995 & 0.8017 & 0.2341 & 3.9835 \\
    & DPCS (ours)         & 30.8417    & 0.9522    & 0.0363    & \textbf{1.5527} \\
    & w/o denoiser (ours) & \textbf{30.8673} & \textbf{0.9529} & \textbf{0.0306} & 1.5539 \\
    & w/o clipping (ours) & 30.0171    & 0.9419    & 0.0563    & 1.6144 \\
    \bottomrule  
    \end{tabular}}
\end{table}

\clearpage
\FloatBarrier

\end{document}